\documentclass[11pt]{article}
\RequirePackage[OT1]{fontenc} \RequirePackage{amsthm,amsmath}
\usepackage{graphics,epsfig,epsf,psfrag,cellspace}
\usepackage{url}
\usepackage{amsmath,amsfonts,amssymb,amsthm}
\usepackage{bbm, bm}
\usepackage{graphicx,lscape,rotate, float, afterpage}
\usepackage{movie15}  
\usepackage{dashrule} 
\usepackage{epstopdf} 
\usepackage{enumerate, framed}
\usepackage{colortbl}
\usepackage{color, xcolor, appendix}
\usepackage[round]{natbib}
\usepackage{movie15}  
\usepackage{dashrule} 
\usepackage{epstopdf} 

\RequirePackage[colorlinks,citecolor=blue,urlcolor=blue]{hyperref}
\usepackage{pstricks,pst-node,pst-tree}  
\usepackage{algorithmic} 
\usepackage[]{algorithm2e}
\usepackage{dsfont,setspace,subcaption}
\newcommand{\algrule}[1][1.pt]{\par\vskip.3\baselineskip\hrule height #1\par\vskip.3\baselineskip}
\makeatother

\usepackage{graphics, hyperref}  

\newtheorem*{remark*}{Remark} 
\theoremstyle{definition}

\makeatletter
\renewcommand*\env@matrix[1][\arraystretch]{%
  \edef\arraystretch{#1}%
  \hskip -\arraycolsep
  \let\@ifnextchar\new@ifnextchar
  \array{*\c@MaxMatrixCols c}}
\makeatother

\newcommand{\z}{\mathbf{z}}

\newcommand{\T}{\mathcal{T}}  

\DeclareMathOperator*{\argmax}{\arg\!\max}

\newcount\colveccount
\newcommand*\colvec[1]{
        \global\colveccount#1
        \begin{pmatrix}
        \colvecnext
}
\def\colvecnext#1{
        #1
        \global\advance\colveccount-1
        \ifnum\colveccount>0
                \\
                \expandafter\colvecnext
        \else
                \end{pmatrix}
        \fi
}

\makeatletter
\newcommand{\Spvek}[2][r]{%
  \gdef\@VORNE{1}
  \left(\hskip-\arraycolsep%
    \begin{array}{#1}\vekSp@lten{#2}\end{array}%
  \hskip-\arraycolsep\right)}

\def\vekSp@lten#1{\xvekSp@lten#1;vekL@stLine;}
\def\vekL@stLine{vekL@stLine}
\def\xvekSp@lten#1;{\def\temp{#1}%
  \ifx\temp\vekL@stLine
  \else
    \ifnum\@VORNE=1\gdef\@VORNE{0}
    \else\@arraycr\fi%
    #1%
    \expandafter\xvekSp@lten
  \fi}
\makeatother

\begin{document}


\title{Enhanced Survival Trees} 
\author{\textbf{Ruiwen Zhou}, \textbf{Ke Xie}, \textbf{Lei Liu} \\
Institute for Informatics, Data Science \& Biostatistics \\
Washington University in St. Louis, MO 63110 \vspace{.1in} \\
\textbf{Zhichen Xu}, \textbf{Jimin Ding} \\
Department of Statistics and Data Science \\
Washington University in St. Louis, MO 63130 \vspace{.1in} \\
and \textbf{Xiaogang Su}\footnote{\url{xsu@utep.edu}} \\
Department of Mathematical Sciences \\
University of Texas, El Paso, TX 79968 \vspace{.1in} 
}

\date{}
\maketitle

\renewcommand{\abstractname}{\large Abstract \vspace{.1in}}
\begin{abstract}
{\normalsize We introduce a new survival tree method for censored failure time data that incorporates three key advancements over traditional approaches. First, we develop a more computationally efficient splitting procedure that effectively mitigates the end-cut preference problem, and we propose an intersected validation strategy to reduce the variable selection bias inherent in greedy searches. Second, we present a novel framework for determining tree structures through fused regularization. In combination with conventional pruning, this approach enables the merging of non-adjacent terminal nodes, producing more parsimonious and interpretable models. Third, we address inference by constructing valid confidence intervals for median survival times within the subgroups identified by the final tree. To achieve this, we apply bootstrap-based bias correction to standard errors. The proposed method is assessed through extensive simulation studies and illustrated with data from the Alzheimer's Disease Neuroimaging Initiative (ADNI) study.
}
\end{abstract}

\noindent%
{\it Keywords:}  Bootstrap bias correction; Censored survival data; Fused regularization; Logrank test; Survival Trees; Variable selection bias.

\section{Introduction}
\label{sec-Introduction}

Survival trees are generally referred to decision trees \citep{Morgan:1963, Breiman:1984} applied to censored survival data. Survival trees have been applied in various biomedical fields where event time is a common outcome, such as the onset of disease, post-treatment relapse, medical diagnosis and prognosis, rehospitalization, and the attainment of developmental milestones.  Interest in survival trees often arises from the need to establish interpretable grouping rules, which help in understanding the scientific structure of the data and in designing future studies. 

Early work on survival trees can be traced back to \cite{Ciampi:1981} and \cite{Gordon:1985}. Numerous approaches have been proposed in the literature. In the conventional CART (Classification and Regression Trees; \citeauthor{Breiman:1984}, \citeyear{Breiman:1984}) framework, the concept of node impurity plays a crucial role in growing and pruning decision trees. Many survival tree methods directly adopt the CART paradigm, utilizing an impurity measure suitable for censored survival data. For instance, \cite{Gordon:1985} used a Wasserstein metric that compares the Kaplan–Meier survival curve at the node with a survival function that has mass at most one finite point. Additionally, \cite{Davis:1989} proposed exponential survival trees based on exponential AFT (accelerated failure time) models, while \cite{Therneau:1990, Keles:2002} applied CART directly to martingale residuals. 

The logrank statistic \citep{Mantel:1966, Peto:1972} is a well-studied two-sample test commonly used in survival analysis. It is intuitive to use the logrank statistic as a splitting criterion in survival trees; see, e.g., \cite{Ciampi:1986} and \cite{Segal:1988}. This approach splits each node by maximizing the difference between its two child nodes, rather than minimizing within-node impurity as in CART. Trees constructed in this manner are referred to as ``trees by goodness of split" by \cite{LeBlanc:1993}, who also developed a split-complexity pruning algorithm, combined with bootstrap bias correction, to prune these trees and determine the final tree model. Survival data typically involve various complexities such as different types of censoring, tied event times, discrete time intervals, time-dependent covariates, clustered subjects, and recurrent events. Research efforts in the literature have sought to extend survival trees to accommodate these complexities. For a more comprehensive survey of survival trees, including an extensive literature list, see \cite{LeBlanc:1995} and \cite{BouHamad:2011}. 

Despite these advancements, several issues persist in the current use of survival trees. First, the greedy search scheme can be time-consuming and is prone to both end-cut preference (ECP) and variable selection bias. Second, practitioners often need to merge or amalgamate the leaves of the final tree models. Third, there is frequent interest in conducting statistical inference at each terminal node of the final tree. While various methods have been proposed to address some of these challenges, they are often heuristic or sub-optimal. Some solutions partially address the problem but introduce new concerns.

In this research, we revisit survival trees based on the logrank test and propose three key enhancements. First, we introduce a computationally efficient data-splitting method to address the end-cut preference issue. This method replaces the threshold function with a smooth sigmoid surrogate (SSS; \citeauthor{Su:2024b}, \citeyear{Su:2024b}) and reformulates the logrank test statistic as a smooth objective function for one-dimensional optimization. We also provide a new approach to mitigate variable selection bias in the standard greedy search. Second, we propose a novel tree modeling approach through fused regularization. This approach, combined with traditional pruning, enables the merging of non-neighboring terminal nodes, resulting in more parsimonious and interpretable tree models. Lastly, we address the inference challenge by constructing valid confidence intervals for the median survival time in each group identified by the final tree model. Our method involves the use of bootstrap bias correction to improve standard error estimates.

The remainder of this article is organized as follows. Section \ref{sec-methods} details the proposed methodological enhancements. Section \ref{sec-simulation} reports simulation studies that evaluate each component of the approach and compare its performance with competing methods. In Section \ref{sec-example}, we demonstrate the application of the proposed method using data from the Alzheimer’s Disease Neuroimaging Initiative (ADNI). Finally, Section \ref{sec-discussion} concludes with a summary of findings and a brief discussion.

\section{Enhanced Survival Trees}
\label{sec-methods}
Consider typical censored survival data that consist of
$\mathcal{D} = \{(T_i,\Delta_i,\mathbf{z}_i) : i=1,...,n\}$, where the observed death time is $T_i=\min(T'_i,C'_i)$ with $(T'_i, C'_i)$ denoting
the death and censoring times for the $i^{th}$ individual;
$\Delta_i=I\{ T'_i \le C'_i \}$ is the survival status indicator;
and $\z_i = \left(z_{ij} \right)_{j=1}^p \in \mathbb{R}^p $ is the
covariate vector associated with subject $i$. For concerns over
identifiability in the ensuing modelling and inference, we assume
that $T'_i$ and $C'_i$ are independent given $\mathbf{z}_i$.

Building a tree model with $\mathcal{D}$ involves three main steps. The first step is to grow a large tree by recursively splitting the data, ensuring important signals are captured. The key element here is the method used for each binary split. The second step is to finalize the tree model, traditionally achieved through a tree pruning algorithm and cross-validation to determine the optimal tree size. The third step is to summarize and present the model. We propose enhancements for each of these steps. 

\subsection{Split with Smooth Sigmoid Surrogate (SSS)}
\label{sec-CutPoint}

The logrank test statistic is commonly used to split survival data \citep{Segal:1988, LeBlanc:1993, Ciampi:1995}. To proceed, consider determining the optimal cutoff point for an ordinal variable \( Z_j \), which we denote as \( Z \) for simplicity. Let \( t_1 < t_2 < \cdots < t_D \) represent the distinct uncensored death times observed within the node \( h \). At each \( t_k \), for \( k = 1, \ldots, D \), a \( 2 \times 2 \) contingency table can be constructed, as outlined below, using a binary split \( I\{z_i \leq c \} \) and  the survival status, where \( z_i \) denotes the \( i \)th observed value of \( Z \) and $c$ represents a cutoff point.

\renewcommand{\tabcolsep}{8.pt}
\renewcommand{\arraystretch}{1.3}
\renewcommand{\baselinestretch}{1.}
\begin{center}
\begin{tabular}{lccc} 
\hline \hline
 & \multicolumn{2}{c}{Child Node of $h$} & \\
\cline{2-3}  &  Left & Right & Total \\
\hline Death &  \cellcolor[gray]{0.9}$d_{kL}$ & $d_k-d_{kL}$ &  \cellcolor[gray]{0.9}$d_k$ \\
Survival & $Y_{kL} - d_{kL}$ & $(Y_k - Y_{kL}) - (d_k-d_{kL})$ &
$Y_k - d_k$ \\ \hline
Total &  \cellcolor[gray]{0.9}$Y_{kL}$ & $Y_k - Y_{kL}$ &  \cellcolor[gray]{0.9}$Y_k$ \\
\hline 
\end{tabular}
\end{center}

In the table above, \( Y_k = \sum_{i=1}^n I\{T_i \geq t_k\} \) represents the total number of subjects at risk at time \( t_k \); \( d_k \) is the total number of deaths at \( t_k \); \( Y_{kL} \) is the number of subjects at risk in the left child node; and \( d_{kL} \) is the number of deaths in the left child node. Other entries can be expressed in terms of \( (Y_k, d_k, Y_{kL}, d_{kL}) \). Given \( (Y_k, d_k, Y_{kL}) \), \( d_{kL} \) follows a hypergeometric distribution with mean \( E_{kL} = Y_{kL} d_k / Y_k \) and variance 
\[
V_{kL} = \frac{d_k (Y_k - d_k) Y_{kL} (Y_k - Y_{kL})}{Y_k^2 (Y_k - 1)},
\]
under the null hypothesis of no difference in survival between the two child nodes. Similar to the Mantel–Haenszel test, the logrank test statistic takes the following form:
\begin{equation}
\label{logrank}
Q(c) = \frac{\left\{ \sum_{k=1}^D w_k (d_{kL} - E_{kL}) \right\}^2}{\sum_{k=1}^D w_k^2 V_{kL}},
\end{equation}
where \( w_k \) is an additional weight associated with \( t_k \). 
Under the condition of no tied death time, it is known that the logrank statistic corresponds to the score test of $H_0:~ \beta=0$ in the \citet{Cox:1972} proportional hazards (PH) model
\begin{equation*} 
\begin{aligned}
\lambda_i(t|\z)=\lambda_0(t)  \, \exp(\beta
\, I\{z_i \leq c \} )
\end{aligned} 
\end{equation*}
where $\lambda_i(t|\z)$ is the hazard function for
subject $i$ and $\lambda_0(t)$ is the baseline hazard.

The optimal cutoff point $c^\star$ is defined as the value that maximizes $Q(c)$ across all permissible cutoff points, expressed as  
$$ c^\star = \argmax_{c} Q(c).$$
Determining $c^\star$ involves evaluating $Q(c)$ at each distinct observed value of $Z$, while satisfying additional constraints, such as ensuring a minimum number of observations per child node (e.g., the \texttt{minbucket} parameter in the \textbf{rpart} package) and a minimum number of deaths in each child node. This discrete optimization, commonly known as a greedy search (GS), becomes computationally intensive when $Z$ has a large number of unique values. To address this challenge, \citet{Su:2024b} introduced the Smooth Sigmoid Surrogate (SSS) approach, offering an efficient alternative for identifying $c^\star$. 

The essential idea of SSS is to replace the threshold function \( I\{z_i \leq c \} \) with a smooth sigmoid function, such as \( \pi\{a (z_i - c)\} \), where \( \pi(x) = \left[ 1 + \exp(-x) \right]^{-1} \) denotes the logistic (or expit) function, and \( a > 0 \) is a shape parameter. This transformation converts the original discrete optimization problem into a smooth optimization problem. To apply SSS, note that \( (d_{kL}, Y_{kL}) \) are the only terms in \( Q(c) \) that involve the cutoff point \( c \). Therefore, we only need to approximate them with \( (\tilde{d}_{kL}, \tilde{Y}_{kL}) \) as follows:
\[
\left\{
\begin{array}{lcl}
Y_{kL}  = \sum_{i=1}^n I \{z_i \leq c \}  \, I\{T_i \geq t_k\} 
& ~~ \longrightarrow ~~ & 
\tilde{Y}_{kL} = \sum_{i=1}^n \pi\{ a (z_i -c)\} \, I\{T_i \geq t_k\} 
\\
d_{kL}  = \sum_{i=1}^n \Delta_i \, I \{z_i \leq c \}  \, I\{T_i \geq t_k\} 
& ~~ \longrightarrow ~~ & 
\tilde{d}_{kL} = \sum_{i=1}^n \Delta_i \, \pi\{ a (z_i -c)\} \, I\{T_i \geq t_k\}.
\end{array}
\right.
\]

As demonstrated through numerical experiments, the performance of SSS is highly robust to the choice of \( a \) \citep{Su:2024b}. For a fixed \( a \), the approximated log-rank test statistic \( \tilde{Q}(c) \) defines a smooth objective function of \( c \), enabling efficient one-dimensional optimization to quickly determine the optimal cutoff point for the covariate \( Z \), i.e.,  
$$ c^\star = \argmax_{c} \tilde{Q}(c). $$ 
In our implementation, both GS and SSS are employed for data splitting, depending on the number of candidate cutoff points for a covariate. Based on the simulation studies in Section \ref{sec-simulation-split}, we recommend using SSS when a covariate has more than 20 cutoff points to evaluate and GS otherwise. Furthermore, SSS can mitigate the end-cut preference (ECP) problem \citep{Breiman:1984} in GS, which refers to the tendency to favor splits that produce small, isolated subsets at the extremes of a feature’s range, even when these splits lack statistical significance or generalizability. We will demonstrate this numerically in Section \ref{sec-simulation-split}. To provide flexibility, our implementation also allows users to opt for either SSS or GS exclusively.

\subsection{Variable Selection Bias}
\label{sec-split}

Suppose that we have identified the optimal cutoff point for each predictor, denoted as $s_j = (Z_j, c^\star_j)$ for $j=1, \ldots, p.$ 
Our next step is to compare these splits, $s_j$, across various predictors to identify the most effective data partition. However, if this comparison is directly based on the maximized logrank statistic, we would encounter the variable selection bias problem \citep{Loh:1991a, Ahn:1994, Loh:2002}, wherein predictors with a greater abundance of potential split points, such as continuous variables, are unduly favored over those with fewer, like binary variables, even if they are not the most informative.

To address this issue, \cite{Loh:2002} proposed first identifying the most important predictor, $Z_{j^\star}$, and then splitting data with its optimal split $s_{j^\star} = (Z_{j^\star}, c^\star_{j^\star})$, where $c^\star_{j^\star}$ denote the optimal cutoff point for $Z_{j^\star}$; similar ideas can be found in other approaches such as \cite{Hothorn:2006}. However, determining the most important predictor remains an equally challenging problem within the recursive partitioning framework. Where various methods exist, it is often unclear whether  and how they ensure the selected predictor is truly the most informative in terms of binary splits. Moreover, although these variable selection techniques are typically designed to assign equal selection probabilities to all predictors in the null scenario where no predictors are associated with the response, they may introduce unintended biases in the non-null scenario. Specifically, as we will demonstrate numerically, these methods tend to reduce the selection probability for predictors with a greater number of distinct values or levels, potentially leading to suboptimal splits.

To mitigate this issue, we propose an intersected validation (IV) approach for variable selection. Our method starts with dividing the data set $\mathcal{D}$ into three mutually exclusive subsets of approximate equal sizes, denoted as $\mathcal{D} = \mathcal{D}_1 + \mathcal{D}_2 + \mathcal{D}_3$. Let $n_k$ denote the sample size of $\mathcal{D}_k$ for $k=1, \ldots, 3.$ We then construct a training sample as  
$$\mathcal{D}'_1 = \mathcal{D}_1 \cup \mathcal{D}^{\mbox{\scriptsize(B)}}_{12},$$ 
where $\mathcal{D}^{\mbox{\scriptsize(B)}}_{12}$ denote a bootstrap sample of $(n_2 + n_3)$ observations drawn from $\mathcal{D}_1 + \mathcal{D}_2.$ Notably, this construction ensures that  $\mathcal{D}'_1$ has the same sample size as the original dataset, $n$. Using $\mathcal{D}'_1$, we determine the optimal cutoff point for each predictor, denoted as $s_j = (Z_j, c^\star_j).$ 

To evaluate the candidate splits $\{s_{j}: j=1, \ldots, p\}$, we construct a validation sample:  
$$\mathcal{D}'_2  = \mathcal{D}_3 \cup \mathcal{D}^{\mbox{\scriptsize(OOB)}}_2 \cup \mathcal{D}^{\mbox{\scriptsize(B)}}_{23},$$ where $\mathcal{D}^{\mbox{\scriptsize (OOB)}}_2$ consists of the out-of-bag (OOB) data from $\mathcal{D}_2$ that were not included in $\mathcal{D}^{(B)}_{12}$, and let $n'_2$ denote its sample size. 
The term $\mathcal{D}^{\mbox{\scriptsize(B)}}_{23}$ represents a bootstrap sample of $(n-n'_2-n_3)$ observations drawn from  $ \mathcal{D}_2 + \mathcal{D}_3$. Using $\mathcal{D}'_2$, we validate each candidate split $s_j$ by recomputing its corresponding logrank test statistic, denoted as  $$Q'(s_j)= Q(c^\star_j)$$ for predictor $Z_j$. The best splitting variable of data $Z_{j^\star}$ is then determined by the maximum validated logrank test statistic, i.e., $ j^\star = \argmax_{j}  Q'(s_j).$

Finally, the optimal cutoff point of $Z_{j^\star}$ is recomputed using the full dataset  $\mathcal{D}$, which yields the best split $s^\star$. 
Let $Q(s^\star)$ denote the corresponding logrank test statistc. According to the best split $s^\star$, the data set $\mathcal{D}$ in the current node $h$ is then partitioned into two child nodes: the left node $h_L$ and the right node $h_R$. The entire IV procedure is outlined in Algorithm \ref{Algorithm-1}. 

The same procedure is applied to partition $h_L$ and $h_R$, till stopping criteria are met. These criteria include a minimum node size and a maximum tree depth, similar to conventional regression trees. In the survival analysis setting, an additional constraint is imposed: each node must contain a sufficient number of uncensored event times to ensure the reliable computation of the logrank test statistic. To maintain comparable proportions of uncensored events between subsamples and the original dataset, the IV procedure employs a stratified sampling approach, using censoring status as the stratification factor. This recursive splitting results in an initial tree, denoted as $\mathcal{T}_0.$

By design, both $\mathcal{D}'_1$ and $\mathcal{D}'_2$ maintain the same sample size $n$ as  the original dataset $\mathcal{D}.$ While the observations in  $\mathcal{D}_3 \cup \mathcal{D}^{\mbox{\scriptsize(OOB)}}_2$ are independent of the training sample $\mathcal{D}'_1$, the resampled set $\mathcal{D}^{\mbox{\scriptsize(B)}}_{23}$ can overlap with $\mathcal{D}^{\mbox{\scriptsize(B)}}_{12}$, meaning that the validation sample $\mathcal{D}'_2$ is not entirely independent of the training sample $\mathcal{D}'_1$. This overlap is the basis for the term intersected validation (IV). The IV design serves two key purposes. First, the validation process provides a more reliable assessment of each predictor’s best split, helping to reduce variable selection bias. Second, maintaining a sample size of $n$ ensures no loss of power in computing the logrank test statistic compared to using the full dataset  $\mathcal{D}.$ As we will demonstrate through simulation studies, the IV approach not only mitigates variable selection bias in the null scenario but also remains effective in non-null scenarios
 
\subsection{Leaf Fusion}
\label{sec-fusion}

Given an initial tree $\mathcal{T}_0$, the goal is to select one of its subtrees as the final tree model. Our discussion is limited to binary trees, and for precise definitions of binary trees, branches, and subtrees, readers are referred to \cite{Breiman:1984}. Let $\mathcal{S}(\mathcal{T}_0)$ denote the set of all subtrees of $\mathcal{T}_0$, which represents the pool of candidate tree models. However, as the size of $\mathcal{T}_0$ increases, the number of subtrees in $\mathcal{S}(\mathcal{T}_0)$ grows exponentially, making it computationally infeasible to evaluate each one. To address this challenge, \cite{Breiman:1984} introduced the cost-complexity pruning algorithm in CART, which follows a bottom-up approach, iteratively removing branches to optimize a cost-complexity performance measure. \cite{LeBlanc:1993} later extended the CART pruning algorithm to survival trees by goodness of split. Specifically, the logrank test statistic is used to maximize the separation between the two child nodes at each split, in contrast to conventional CART, which minimizes within-node impurity. This specialized pruning technique for survival trees is termed the split-complexity algorithm. The pruning algorithm narrows down the vast number of candidate models in $\mathcal{S}(\mathcal{T}_0)$ into a nested sequence of optimally pruned subtrees, from which the final tree model can then be determined via validation. 

In this study, we expand the set of candidate tree models, $\mathcal{S}(\mathcal{T}_0)$, by allowing node fusion in any subtree of $\mathcal{T}_0$. This enhances flexibility and increases the likelihood of identifying an improved final tree model. Node fusion or amalgamation is not a new concept; it has been applied \textit{post hoc} to a final tree model \citep{Ciampi:1988, Ciampi:1989, Fan:2006} by  performing pairwise comparisons among terminal nodes and iteratively merging those with the smallest, statistically insignificant differences. This approach can produce a more parsimonious model with improved interpretability. However, these methods are typically \textit{post hoc} and lack cross-validation. Recently, \cite{Su:2024a} introduced TreeFuL (Trees with Fused Leaves), a method that achieves node fusion through fused regularization, providing an alternative to the traditional pruning algorithm. The initial tree structure, $\mathcal{T}_0$, naturally defines a grouping of observations, with each observation assigned to exactly one terminal node. If these terminal nodes can be further grouped, a rapid top-down shearing procedure can be applied to obtain a subtree. Starting from the root node, this shearing process follows a simple rule: prune any node whose descendants all belong to the same group. In this work, we extend the TreeFuL approach to survival trees, referred to as SurvTreeFuL.

\subsubsection{Fused Regularization}
\label{sec-FusedRegularization}
Let $\widetilde{\mathcal{T}}_0 = \{h_1, \ldots, h_K\}$ represent the set of terminal nodes in a tree $\mathcal{T}_0$, and let $|\cdot|$ denote cardinality, so that $K = |\widetilde{\mathcal{T}}_0|$ is the number of terminal nodes in $\mathcal{T}_0$. Define the dummy vector $\bm{x}_i = \left(x_{i1}, \ldots, x_{i(K-1)} \right)^T \in \mathbb{R}^{K-1}$, where each element is given by  
$$
x_{ik} =
\begin{cases}
1, & \text{if the } i\text{th observation falls into terminal node } h_k \text{ of } \mathcal{T}_0, \\
0, & \text{otherwise},
\end{cases}
$$
for $i = 1, \ldots, n$ and $k = 1, \ldots, K-1$. The $K$th terminal node is omitted as the reference category. Consider the Cox proportional hazards (PH) model:  
\begin{equation}
\label{cox-model}
\lambda_i(t|\bm{x}_i) = \lambda_0(t) \exp \left( \sum_{k=1}^{K-1} \beta_k x_{ik}  \right) = \lambda_0(t) \exp \left( \bm{x}^T_i \bm{\beta}  \right), 
\end{equation}
with $\bm{\beta} = (\beta_1, \ldots, \beta_{K-1})^T \in \mathbb{R}^{K-1}$. By fitting this model, the terminal nodes $h_k$ can be sorted and relabeled based on the maximum partial likelihood estimates (MPLE) $\hat{\beta}_k$, arranged in ascending order such that  
$$ \beta_0=\hat{\beta}_0=0 \leq \hat{\beta}_1 \leq  \hat{\beta}_2 \leq \cdots \leq \hat{\beta}_{K-1}, $$
where $\beta_0=\hat{\beta}_0=0$ corresponds to the reference node with the smallest hazard rate.  With a slight abuse of notation, we use the same index $k$ to denote the sorted leaves. Alternatively, if available, the median survival time within each $h_k$ can be used for sorting.  

The fused regularization \citep{Tibshirani:2005} can be expressed as the following optimization problem:
\begin{equation}
\label{optim1}
\min_{\bm{\beta}} ~ - \frac{2}{n} L(\bm{\beta}) + \lambda \sum_{k=1}^{K-1} w_k \left| \beta_k - \beta_{k-1} \right|,
\end{equation}
where 
\begin{equation}
\label{partial-loglik}
L(\bm{\beta}) = \sum_{i=1}^n \Delta_i \left[ \bm{x}_i^T \bm{\beta} 
- \log \sum_{i'=1}^n \left\{ I(T_{i'} \geq T_i) \, \exp\left( \bm{x}_{i'}^T \bm{\beta} \right) \right\} \right]
\end{equation}
is the partial log-likelihood associated with model~(\ref{cox-model}); 
$\lambda \geq 0$ is a tuning parameter; and $w_k = 1 / |\hat{\beta}_k - \hat{\beta}_{k-1}|$ corresponds to the adaptive LASSO weights \citep{Zou:2006}. 
For notational convenience, we set $\beta_0 = \hat{\beta}_0 = 0$. This implies that $\beta_0$ is not estimated, and the first penalty term in~(\ref{optim1}) simplifies to $|\beta_1 - \beta_0| = |\beta_1|$, with weight $w_1 = 1 / |\hat{\beta}_1|$. This encourages the fusion of other nodes with the reference node that has the lowest hazard rate. The remaining penalty terms, of the form $w_k \left| \beta_k - \beta_{k-1} \right|$, promote the fusion of neighboring nodes. While alternative penalty functions may be used, the adaptive LASSO formulation enables us to leverage the well-established implementation available in the \textbf{glmnet} R package \citep{Friedman:2010}.

To solve (\ref{optim1}), we introduce the matrix
\begin{equation}
\label{B}
\bm{B} = 
\begin{bmatrix}[1.0]
1 & 0 & 0 & \cdots & 0 & 0 \\
-1 & 1 & 0 & \cdots & 0 & 0 \\
0 & -1 & 1 & \cdots & 0 & 0 \\
\vdots & \vdots & \vdots & \cdots & \vdots & \vdots \\
0 & 0 & 0 & \cdots & -1 & 1 
\end{bmatrix}  \in \mathbb{R}^{(K-1) \times (K-1)} 
\end{equation}
and the diagonal matrix  $\bm{W} = \mbox{diag}\left( w_k \right) \in \mathbb{R}^{(K-1) \times (K-1)}.$ Define $\bm{\gamma} = \bm{W} \bm{B} \bm{\beta}.$ It follows that $ \bm{\beta} = \bm{B}^{-1} \bm{W}^{-1} \bm{\gamma}$, where 
$$ \bm{W}^{-1} = \mbox{diag}\left( 1/w_k \right) = \mbox{diag}\left( |\hat{\beta}_k - \hat{\beta}_{k-1}| \right) $$
and 
\begin{equation}
\label{B-inverse}
\bm{B}^{-1} = 
\begin{bmatrix}[1.0]
1 & 0 & 0 & \cdots & 0 & 0 \\
1 & 1 & 0 & \cdots & 0 & 0 \\
1 & 1 & 1 & \cdots & 0 & 0 \\
\vdots & \vdots & \vdots & \cdots & \vdots & \vdots \\
1 & 1 & 1 & \cdots & 1 & 1 
\end{bmatrix}  
\end{equation}
is a lower-triangular matrix with all lower-triangular entries equal to 1.
Define the transformed covariate vector $\bm{x'}_i = \bm{W}^{-1} \bm{B}^{-1} \bm{x}_i$. In terms of the design matrices $\bm{X} = \left( \bm{x}_i^T \right)$—where $\bm{x}_i^T$ is the $i$th row vector—and $\bm{X}' = \left( \bm{x'}^T_i \right)$, this transformation amounts to
$ \bm{X}' = \bm{X} \bm{B}^{-1} \bm{W}^{-1}. $

The model in~(\ref{cox-model}) can then be reformulated as
\begin{equation}
\label{cox-model2}
\lambda_i(t \mid \bm{x}'_i) = \lambda_0(t) \exp \left( \bm{x}'_i{}^T \bm{\gamma} \right).
\end{equation}
Furthermore, the optimization problem in~(\ref{optim1}) can be equivalently rewritten as a standard LASSO problem for Cox proportional hazards (PH) models:
\begin{equation}
\label{optim2}
\min_{\bm{\gamma}}~ - \frac{2}{n} L(\bm{\gamma}) + \lambda \sum_{k=1}^{K-1} |\gamma_k|,
\end{equation}
where $L(\bm{\gamma})$ denotes the partial log-likelihood associated with model~(\ref{cox-model2}). The solution path $\{ \widetilde{\bm{\gamma}}(\lambda) : \lambda \geq 0 \}$ for problem~(\ref{optim2}) can be efficiently obtained using the \textbf{glmnet} R package \citep{Simon:2011}. The corresponding regularization path $\{ \widetilde{\bm{\beta}}(\lambda) : \lambda \geq 0 \}$ for problem~(\ref{optim1}) is then recovered via the transformation
$$
\widetilde{\bm{\beta}}(\lambda) = \bm{B}^{-1} \bm{W}^{-1} \widetilde{\bm{\gamma}}(\lambda).
$$
The solution path is a piecewise linear function of $\lambda$. Let $\lambda_m$ denote the value of $\lambda$ at which a change in slope occurs, and let $M$ be the total number of such change points. The solution path can then be represented as
\[
\left\{ \widetilde{\bm{\beta}}_m = \widetilde{\bm{\beta}}(\lambda_m) : m = 1, \ldots, M \right\}.
\]
When $\lambda = \lambda_1 = 0$, the solution $\widetilde{\bm{\beta}}(0)$ corresponds to the maximum partial likelihood estimator (MPLE), denoted by $\widehat{\bm{\beta}}$. As $\lambda$ increases, each solution $\widetilde{\bm{\beta}}(\lambda_m)$ defines a distinct grouping structure for the terminal nodes in $\mathcal{T}_0$ by merging them in different ways. At the final value $\lambda = \lambda_M$, all terminal nodes are merged into a single group.

\subsubsection{Tuning Parameter Selection}
\label{sec-tuning}
Our next objective is to determine the optimal tuning parameter $\lambda^\star$ through validation. Two commonly used approaches are the test sample method and $V$-fold cross-validation. A natural performance metric for this purpose is the validated deviance \citep{LeBlanc:1992}, defined as
\begin{equation}
\label{deviance}
D(\mathcal{D}, \mathcal{D}') = 2 \sum_{i \in \mathcal{D}_2} \left[ \hat{\Lambda}_0(T_i) \exp \left( \bm{x}_i^T \widehat{\bm{\beta}} \right) 
- \Delta_i \left\{ 1 + \bm{x}_i^T \widehat{\bm{\beta}} + \log \hat{\Lambda}_0(T_i) \right\} \right],
\end{equation}
where $\hat{\Lambda}_0(\cdot)$ denotes the \cite{Breslow:1972} estimator of the cumulative hazard function. In this definition, both $\hat{\Lambda}_0(\cdot)$ and $\widehat{\bm{\beta}}$ are computed using the training data $\mathcal{D}$, while the deviance is evaluated on the validation set $\mathcal{D}'$, consisting of observations $\{(T_i, \Delta_i, \bm{x}_i)\}$. 

In the case of sufficiently large data, a test-sample approach can be applied. We randomly partition the data into a training set $\mathcal{D}$ and a testing set $\mathcal{D}'$. Based on the training set $\mathcal{D}$, an initial tree $\mathcal{T}_0$ is constructed. Applying fused regularization yields the set 
\[
\left\{ (\lambda_m, \widetilde{\bm{\beta}}_m, \widehat{\bm{\beta}}_m, \hat{\Lambda}_{0m}(\cdot)) : m = 1, \ldots, M \right\},
\]
where $\widehat{\bm{\beta}}_m$ denotes the `relaxed' LASSO estimator, i.e., the MPLE obtained without penalization, corresponding to the sparsity or fusion pattern given by $\widetilde{\bm{\beta}}_m$, and $\hat{\Lambda}_{0m}(\cdot)$ is the estimated cumulative baseline hazard function based on $\widehat{\bm{\beta}}_m$. Next, the testing set $\mathcal{D}'$ is passed through the initial tree $\mathcal{T}_0$, and the validated deviance $D_m$ is computed as
$$ D_m = D_m(\mathcal{D}, \mathcal{D}') = 2 \sum_{i \in \mathcal{D}'} \left[ \hat{\Lambda}_{0m}(T_i) \exp \left( \bm{x}_i^T \widehat{\bm{\beta}}_m \right) 
- \Delta_i \left\{ 1 + \bm{x}_i^T \widehat{\bm{\beta}}_m + \log \hat{\Lambda}_{0m}(T_i) \right\} \right]. $$

In situations where data are limited, $V$-fold cross-validation provides an effective strategy for model selection. The procedure begins by constructing an initial tree $\mathcal{T}_0$ using the entire dataset $\mathcal{D}$. Fused regularization is then applied to obtain a regularization path
$ \left\{ (\lambda_m, \widetilde{\bm{\beta}}_m) : m = 1, \ldots, M \right\},$
which yields a sequence of candidate models indexed by distinct tuning parameter values $\lambda_m$.

To perform cross-validation, the dataset $\mathcal{D}$ is randomly partitioned into $V$ approximately equal-sized folds, denoted by $\left\{ \mathcal{D}_v : v = 1, \ldots, V \right\}$. For each fold $v$, let $\mathcal{D}^{(v)} = \mathcal{D} \setminus \mathcal{D}_v$ denote the training data excluding the $v$-th fold. Based on $\mathcal{D}^{(v)}$, a new tree $\mathcal{T}_v$ is grown, and fused regularization is applied using the same sequence of tuning parameters $\{\lambda_m : m = 1, \ldots, M\}$ identified from the full data. This yields
$$
\left\{ \left(\lambda_m, \widetilde{\bm{\beta}}_m^{(v)}, \widehat{\bm{\beta}}_m^{(v)}, \hat{\Lambda}_{0m}^{(v)}(\cdot) \right) : m = 1, \ldots, M \right\},
$$
where $\widehat{\bm{\beta}}_m^{(v)}$ denotes the relaxed estimator associated with $\widetilde{\bm{\beta}}_m^{(v)}$, and $\hat{\Lambda}_{0m}^{(v)}(\cdot)$ is the corresponding cumulative baseline hazard function.

The hold-out fold $\mathcal{D}_v$ is then passed through $\mathcal{T}_v$, and the validation deviance is computed for each $m$ as
$$ D_{mv} = 2 \sum_{i \in \mathcal{D}_v} \left[ \hat{\Lambda}_{0m}^{(v)}(T_i) \exp \left( \bm{x}_i^T \widehat{\bm{\beta}}_m^{(v)} \right) 
- \Delta_i \left\{ 1 + \bm{x}_i^T \widehat{\bm{\beta}}_m^{(v)} + \log \hat{\Lambda}_{0m}^{(v)}(T_i) \right\} \right]. $$
This process is repeated for each of the $V$ folds. Since deviance is additive across independent data subsets, the overall cross-validated deviance for each $\lambda_m$ can be computed as
$$ D_m = \sum_{v=1}^V D_{mv}.$$

The complete cross-validation (CV) procedure is described in Algorithm~\ref{Algorithm-2}. Several important remarks are in order. First, the standard $V$-fold CV approach commonly used in conventional regularization frameworks cannot be directly applied in our context. This limitation arises because the initial tree $\mathcal{T}_0$ is itself learned from the data. As such, the validation fold $\mathcal{D}_v$ must be used to assess the entire modeling pipeline—beginning with the construction of the tree $\mathcal{T}_v$ and followed by the application of fused regularization—based solely on the corresponding training fold $\mathcal{D}^{(v)}$. Second, the inclusion of the scaling factor $2/n$ in the objective function~\eqref{optim1} is critical for maintaining the comparability of the tuning parameter $\lambda$ across datasets of varying sizes. This normalization allows a fixed set of tuning parameters, $\{\lambda_m : m = 1, \ldots, M\}$, to be applied uniformly across all training subsets $\mathcal{D}^{(v)}$ during cross-validation. Third, we recommend using stratified sampling based on censoring status when partitioning the data. This helps maintain a consistent censoring rate across different folds or resampled datasets, thereby improving the stability and reliability of the model evaluation process.

In both the test sample and $V$-fold cross-validation approaches, the optimal tuning parameter $\lambda^\star$ is selected by minimizing the validated deviance:
$$ \lambda^\star = \lambda_{m^\star}, \quad \text{where} \quad m^\star = \arg\min_m D_m. $$
As an alternative, information-theoretic model selection criteria such as the Akaike Information Criterion (AIC; \citealp{Akaike:1974}) and the Bayesian Information Criterion (BIC; \citealp{Schwarz:1978}) can be employed in a heuristic manner \citep{LeBlanc:1993, Su:2004}. These criteria augment the deviance with a penalty term for model complexity and take the general form
$$ D_m + \lambda_0 K_m, $$
where $K_m$ denotes the number of distinct groups formed in the fused model corresponding to $\lambda_m$, and $\lambda_0$ is a penalty constant: $\lambda_0 = 2$ for AIC and $\lambda_0 = \log(n')$ for BIC. Here, $n'$ refers to the number of uncensored events in the validation set $\mathcal{D}'$ for the test sample approach or in the full dataset $\mathcal{D}$ for the cross-validation setting. 

\subsubsection{Coloring and Shearing}
\label{sec-shear}

Once the optimal tuning parameter $\lambda^\star$ is selected, the corresponding fused estimator $\widetilde{\bm{\beta}}^\star = \widetilde{\bm{\beta}}(\lambda^\star)$ is obtained. The fusion pattern encoded in $\widetilde{\bm{\beta}}^\star$ determines how the terminal nodes (leaves) of the initial tree $\mathcal{T}_0$ should be grouped. To visually represent this grouping, we assign a distinct color to each fused group, such that all leaves belonging to the same group are labeled with the same color.

Using this color-based grouping, we then simplify $\mathcal{T}_0$ through a procedure we term \emph{shearing}. This terminology is intentionally chosen to distinguish it from the standard notion of \emph{pruning} in the CART literature. Shearing is a top-down operation that traverses the tree from the root, applying a straightforward rule: an internal node is pruned if all of its descendant leaves share the same color. More formally, for each internal (non-terminal) node $h$ in $\mathcal{T}_0$, we examine the leaves in its subtree; if they are all assigned the same color, the entire subtree rooted at $h$ is removed, and $h$ is converted into a new terminal node, inheriting the common color of its pruned descendants.

The outcome of this shearing process is a simplified tree, denoted by $\mathcal{T}^\star$, which represents the minimal tree structure consistent with the fused model. Importantly, $\mathcal{T}^\star$ may contain non-adjacent leaves that share the same color; in such cases, these leaves are treated as a single group by the SurvTreeFuL method. Consequently, the final SurvTreeFuL model is fully characterized by the sheared tree $\mathcal{T}^\star$ together with the color-based grouping of its terminal nodes.

An illustration of this coloring and shearing procedure is presented in Figure~I of the Supplementary Materials. It demonstrates how shearing can be performed efficiently when the grouping structure of the terminal nodes is known. 

\subsection{Valid Inference}
\label{sec-BBC}
Finally, we provide descriptive summaries for each group identified by the final tree model $\mathcal{T}^\star$. In the R package \textbf{party} \citep{Hothorn:2006, Zeileis:2008}, group summaries are based on median survival times, whereas \textbf{rpart} \citep{Therneau:1990} reports hazard ratios using the root node as the baseline. Since our method employs the logrank test statistic for splitting, it is natural to summarize the resulting groups using Cox model coefficients, which correspond to the logarithm of the hazard ratio. However, due to the highly adaptive nature of recursive partitioning, it is important to consider whether these estimates, and their associated standard errors, are subject to systematic bias.

We examine this issue through simulation. Panels (a) and (b) in Figure~\ref{fig04-BBC} illustrate results based on data generated from a tree model (Model C in Table~\ref{tbl-SimModels}). Using the groupings from the final tree, a Cox proportional hazards model is fit with the group exhibiting the lowest hazard designated as the baseline. Estimated coefficients and their corresponding standard errors (SEs) are recorded, where SEs are rescaled to standard deviations (SDs) by multiplying by $\sqrt{n}$. Since the true coefficient values are generally unknown, we generate a large test sample ($n' = 100{,}000$) and recompute these quantities for comparison. As shown in Figure~\ref{fig04-BBC}(a) and (b), both the estimated coefficients and SDs based on the training data tend to be inflated, exhibiting a consistent upward bias relative to the test sample estimates, which serve as a benchmark. This pattern of overestimation is also observed under other underlying models (see Supplementary Figure~II), which is unsurprising given that the tree construction maximizes between-node hazard differences, thereby exaggerating hazard ratios in the final model. The larger SD may be attributed to the fact that variables with greater magnitude tend to exhibit greater variability. Interestingly, similar experiments using alternative summary measures, such as median survival time and constant hazard rate, show much closer agreement between estimates obtained from the training and test data. This observation suggests that the adaptive nature of recursive partitioning tends to introduce bias primarily in quantities directly involved in the splitting criterion. A similar perspective was previously noted by \citet{Loh:1991a, Loh:2002}.

The above observation motivates us to address the bias in estimating $\beta$ and its associated standard deviation (SD). A widely used approach for bias correction is the bootstrap method \citep{Efron:1993}. In this framework, the estimate from each bootstrap sample is compared to the estimate from the original training data, and the average of these differences provides an estimate of the bias. However, applying this method to tree-based models presents a challenge due to their inherent instability: the tree structures generated from different bootstrap samples often vary substantially from one another, and from the final tree built on the original dataset. To address this, we exploit a key property of tree models that each tree partitions the dataset into distinct groups. Given two tree structures, observations that belong to a group in one tree may be distributed across different groups in the other. This mapping allows us to estimate the bias as a weighted average across group comparisons. Leveraging this idea, we propose a modified bootstrap bias correction procedure, focusing on the estimation of $\beta$. A similar approach applies to correcting its SD.

Suppose the final tree model $\mathcal{T}^\star$ is fitted using the dataset $\mathcal{D}$, and partitions the data into $K^\star$ groups via node fusion. For each group $k = 1, \ldots, K^\star$, let $\hat{\beta}_k$ denote the estimated coefficient. By convention, we set $\hat{\beta}_1 = 0$ for the baseline group (i.e., the group with the lowest hazard). Let $\bm{g} = (g_1, \ldots, g_n)^\top \in \mathbb{R}^n$ denote the group membership vector such that $g_i = k$ if the $i$th observation in $\mathcal{D}$ belongs to group $k$.

We draw $B$ bootstrap samples $\{ \mathcal{D}_b : b = 1, \ldots, B \}$. For each bootstrap sample $\mathcal{D}_b$, we grow a tree $\mathcal{T}_b$ of the same depth $d^\star$ as $\mathcal{T}^\star$, and fuse its leaves to form $K_b$ groups. Ideally, $K_b = K^\star$, but if that is not feasible, we choose the smallest possible $K_b > K^\star$. Although $\mathcal{T}_b$ could be constructed using a test set or cross-validation as done for $\mathcal{T}^\star$, this alternative construction substantially reduces computational burden. 

Let $\hat{\beta}_{k'}(\mathcal{D}', \mathcal{T})$ denote the estimate of $\beta_{k'}$ based on dataset $\mathcal{D}'$ and tree $\mathcal{T}$. By default, $$\hat{\beta}_k = \hat{\beta}_k(\mathcal{D}, \mathcal{T}^\star).$$ For each bootstrap sample $\mathcal{D}_b$ and tree $\mathcal{T}_b$, we compute $\hat{\beta}_{k'}(\mathcal{D}_b, \mathcal{T}_b)$ and $\hat{\beta}_{k'}(\mathcal{D}, \mathcal{T}_b)$ for $k' = 1, \ldots, K_b$. The corresponding bootstrap bias estimate is given by
\[
\tau_{bk'} = \hat{\beta}_{k'}(\mathcal{D}_b, \mathcal{T}_b) - \hat{\beta}_{k'}(\mathcal{D}, \mathcal{T}_b).
\]

Our goal, however, is to estimate the bias $\tau_k$ for each $\hat{\beta}_k$, where $k = 1, \ldots, K^\star$. We achieve this by computing a weighted average of the $\tau_{bk'}$ values, based on how the observations in group $k$ (from $\mathcal{T}^\star$) are distributed across the $K_b$ groups (from $\mathcal{T}_b$). Let $\bm{g}_b \in \mathbb{R}^n$ denote the group membership vector for $\mathcal{T}_b$, assigning each observation in $\mathcal{D}$ to one of the $K_b$ groups. The pair $(\bm{g}, \bm{g}_b)$ defines a $K^\star \times K_b$ contingency table with cell counts $\{n_{kk'}\}$. Let $p_{kk'} = n_{kk'}/n_{k\cdot}$ be the proportion of observations in group $k$ (from $\mathcal{T}^\star$) that fall into group $k'$ (from $\mathcal{T}_b$), where $n_{k\cdot} = \sum_{k'} n_{kk'}$ is the row total. Then, the bias estimate for $\hat{\beta}_k$ from the $b$th bootstrap sample is
$ \sum_{k'=1}^{K_b} p_{kk'} \tau_{bk'}.$
Averaging over $B$ bootstrap samples yields the final bias-corrected estimate for $\beta_k$:
\begin{align}
\hat{\beta}_k &:= \hat{\beta}_k + \frac{1}{B} \sum_{b=1}^B \sum_{k'=1}^{K_b} p_{kk'} \tau_{bk'} \nonumber \\
&:= \hat{\beta}_k + \frac{1}{B} \sum_{b=1}^B \sum_{k'=1}^{K_b} \frac{n_{kk'}}{n_{k\cdot}} \left[ \hat{\beta}_{k'}(\mathcal{D}_b, \mathcal{T}_b) - \hat{\beta}_{k'}(\mathcal{D}, \mathcal{T}_b) \right].
\label{eqn-BBC}
\end{align}

The complete procedure is outlined in Algorithm~\ref{Algorithm-3}. For bias correction purposes, we generally recommend selecting $B$ within the range $20 \leq B \leq 50$ \citep{Efron:1993, LeBlanc:1993}.  The same method can be applied to correct bias in estimating the standard deviation (SD) of $\hat{\beta}_k$. The standard error (SE) can then be obtained by scaling the SD by a factor of $1/\sqrt{n}$. Combined with the bias-corrected estimates $\tilde{\beta}_k$, this enables valid statistical inference, including confidence interval (CI) construction.

Regarding statistical inference for decision trees, \cite{Loh:2018} proposed a bootstrap calibration approach \citep{Loh:1991b} for constructing confidence intervals in tree-structured subgroup analyses. More recently, \cite{Neufeld:2022} extended selective inference to CART models. To the best of our knowledge, this work represents the first  approach to valid statistical inference with survival trees.

\section{Simulation Studies}
\label{sec-simulation}
This section presents simulation studies conducted to evaluate each step of SurvTreeFuL and compare the proposed approaches with competitive methods. 

\subsection{Optimal Cutoff Point}
\label{sec-simulation-split}

We first compare the Smooth Sigmoid Surrogate (SSS) approach, introduced in Section~\ref{sec-CutPoint}, with the conventional Greedy Search (GS) method for identifying the optimal cutoff point of a continuous predictor. The data are generated from the following model:
\begin{equation}
\label{model-split}
\lambda(t) = \exp \left[\beta_0 + \beta_1 \, I(z \leq c_0) \right],
\end{equation}
where the covariate $Z$ follows a uniform distribution on $[0, 1]$, the true cutoff point is set at $c_0 = 0.5$, and the coefficients are specified as $\beta_0 = 1$ and $\beta_1 = -1$. Although we consider a range of sample sizes and censoring rates, we report the results here for the case with sample size $n = 200$ and a censoring rate of 50\%. For each simulated dataset, both GS and SSS are applied to estimate the optimal cutoff point $\hat{c}$. A key parameter in SSS is the shape parameter $a > 0$; to investigate its influence, we vary $a$ over the set $\{5, 10, 15, 20, \ldots, 100\}$. A total of 1,000 simulation runs are conducted for each model configuration. 

\begin{figure}[H]
     \centering
     \begin{subfigure}[b]{0.47\textwidth}
         \centering
         \includegraphics[scale=.7, angle=270, width=\textwidth]{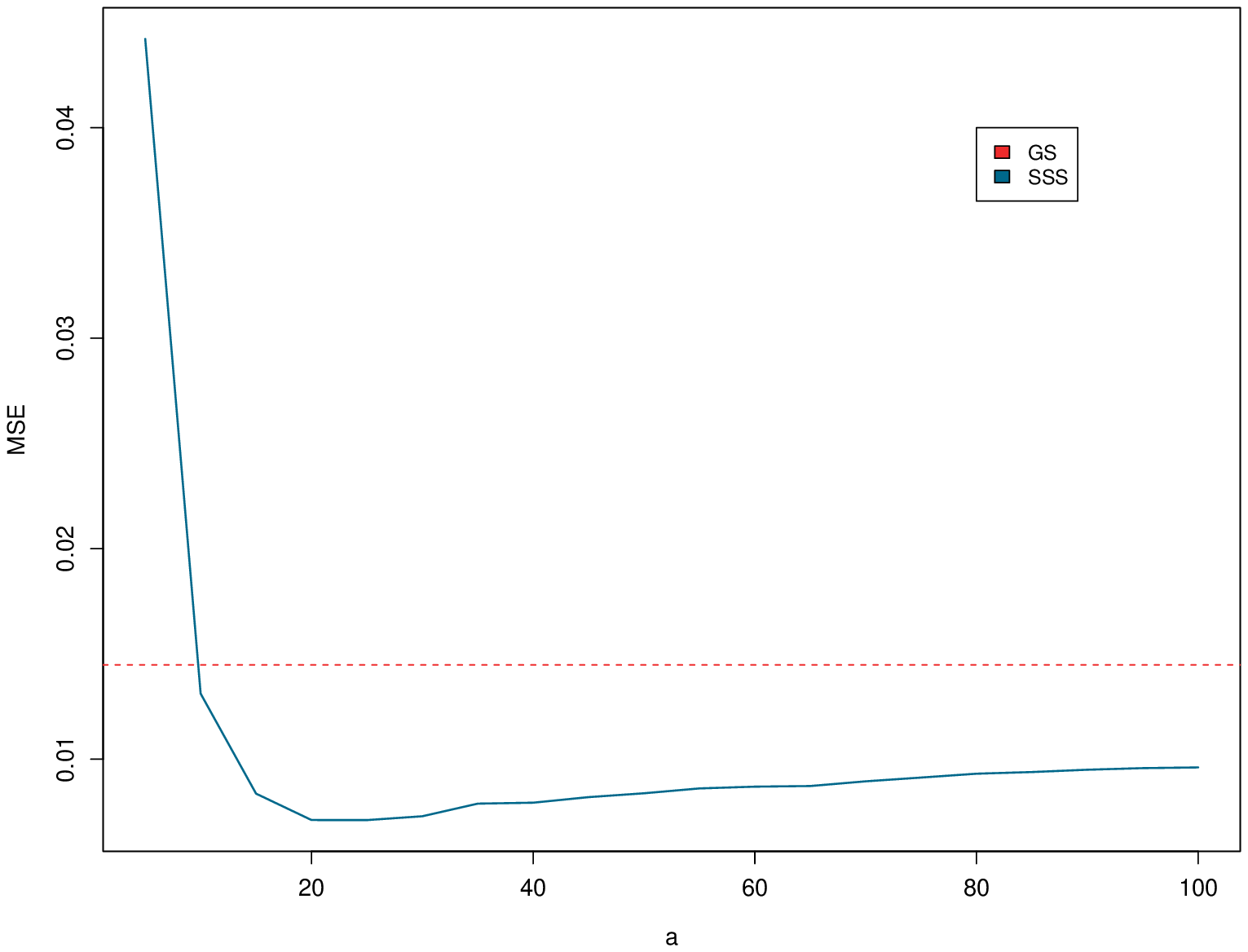}
         \caption{MSE}
         \label{fig01a}
     \end{subfigure}
     \hfill
     \begin{subfigure}[b]{0.47\textwidth}
         \centering
         \includegraphics[scale=.7,angle=270, width=\textwidth]{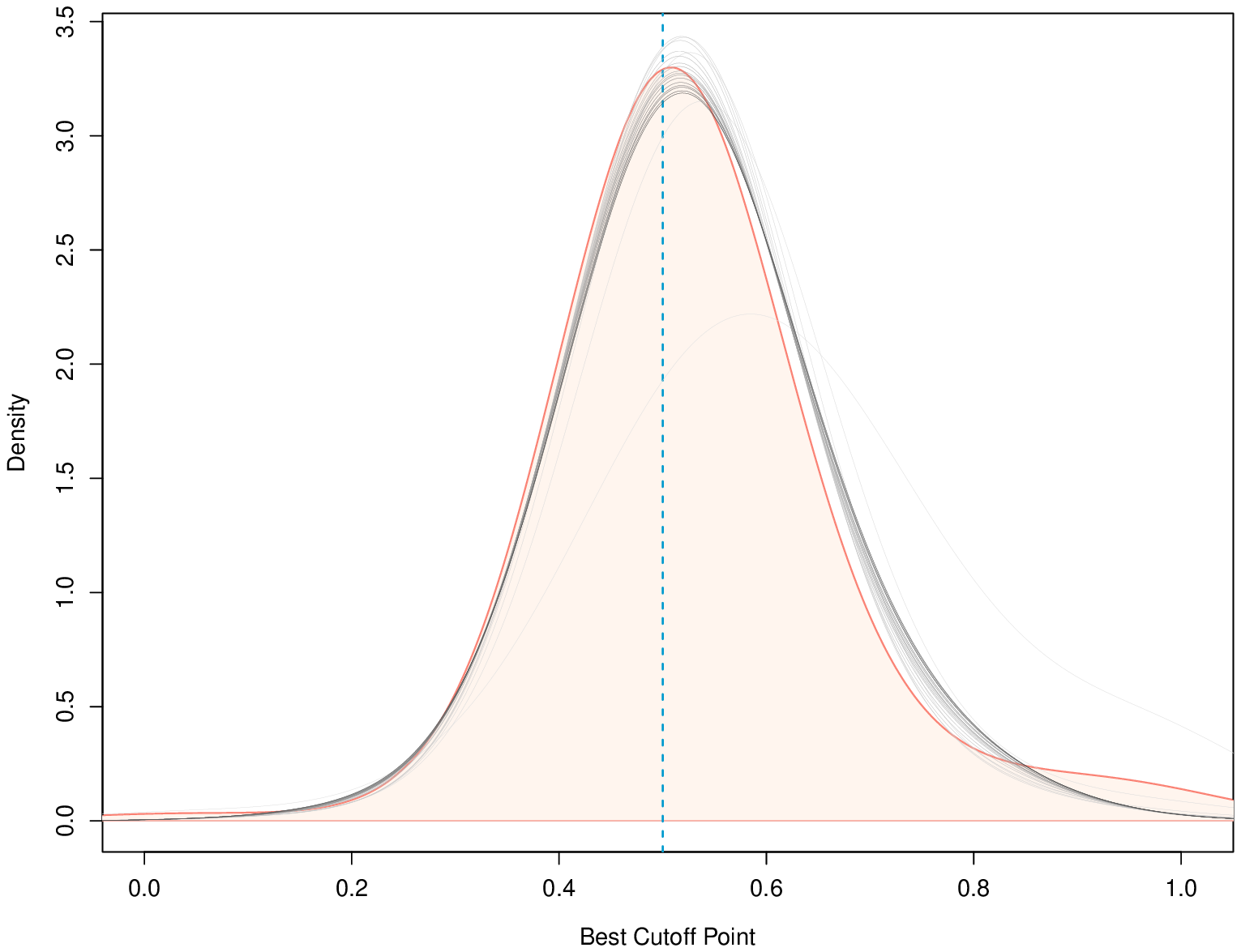}
        \caption{Density Plots}
         \label{fig01b}
     \end{subfigure}
     
\vspace{0.3in}

     \begin{subfigure}[b]{0.47\textwidth}
         \centering
         \includegraphics[scale=.7,angle=270, width=\textwidth]{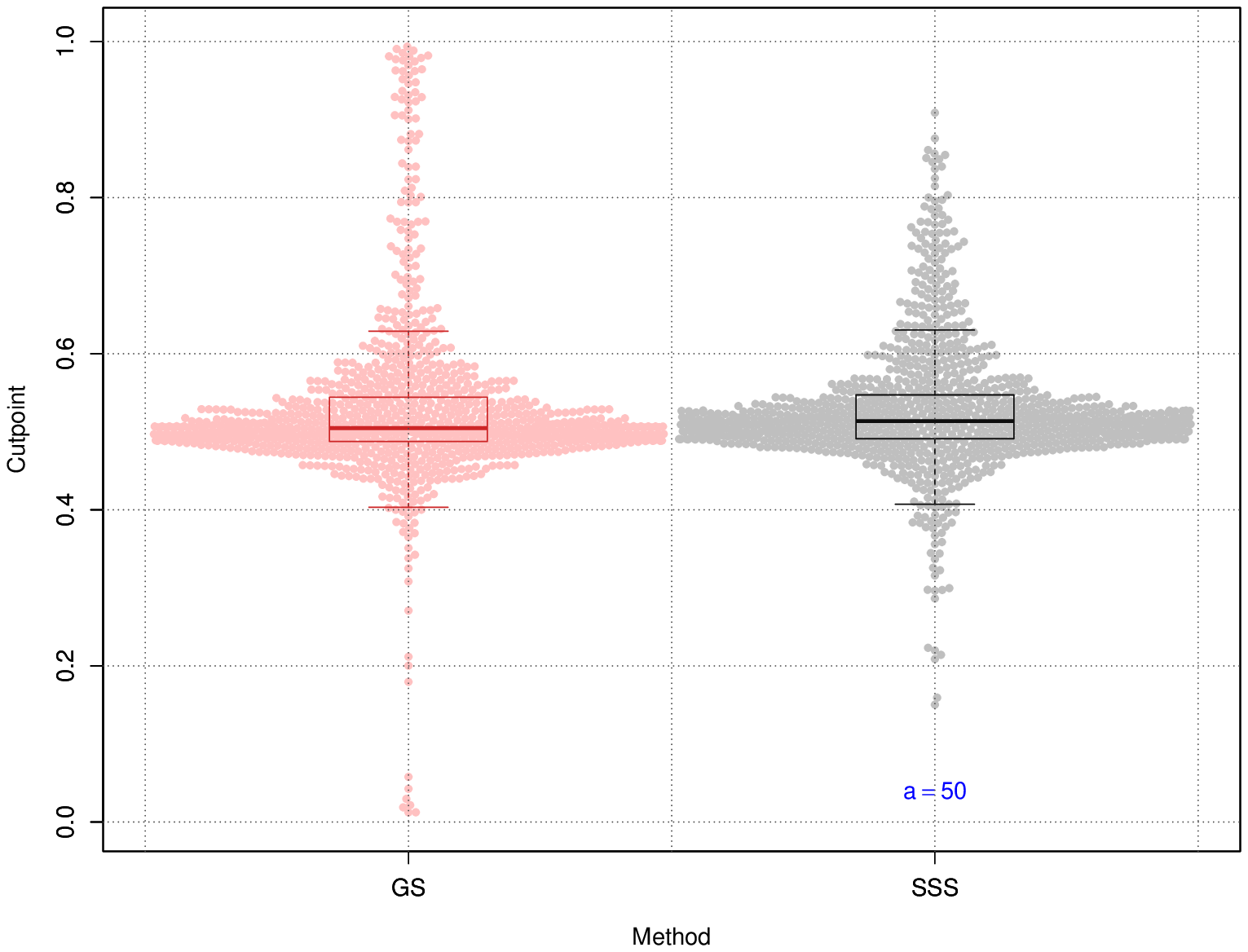}
         \caption{Bee Swarm Plot and Boxplot}
         \label{fig01c}
     \end{subfigure}
 \hfill
     \begin{subfigure}[b]{0.47\textwidth}
         \centering
         \includegraphics[scale=.7,angle=270, width=\textwidth]{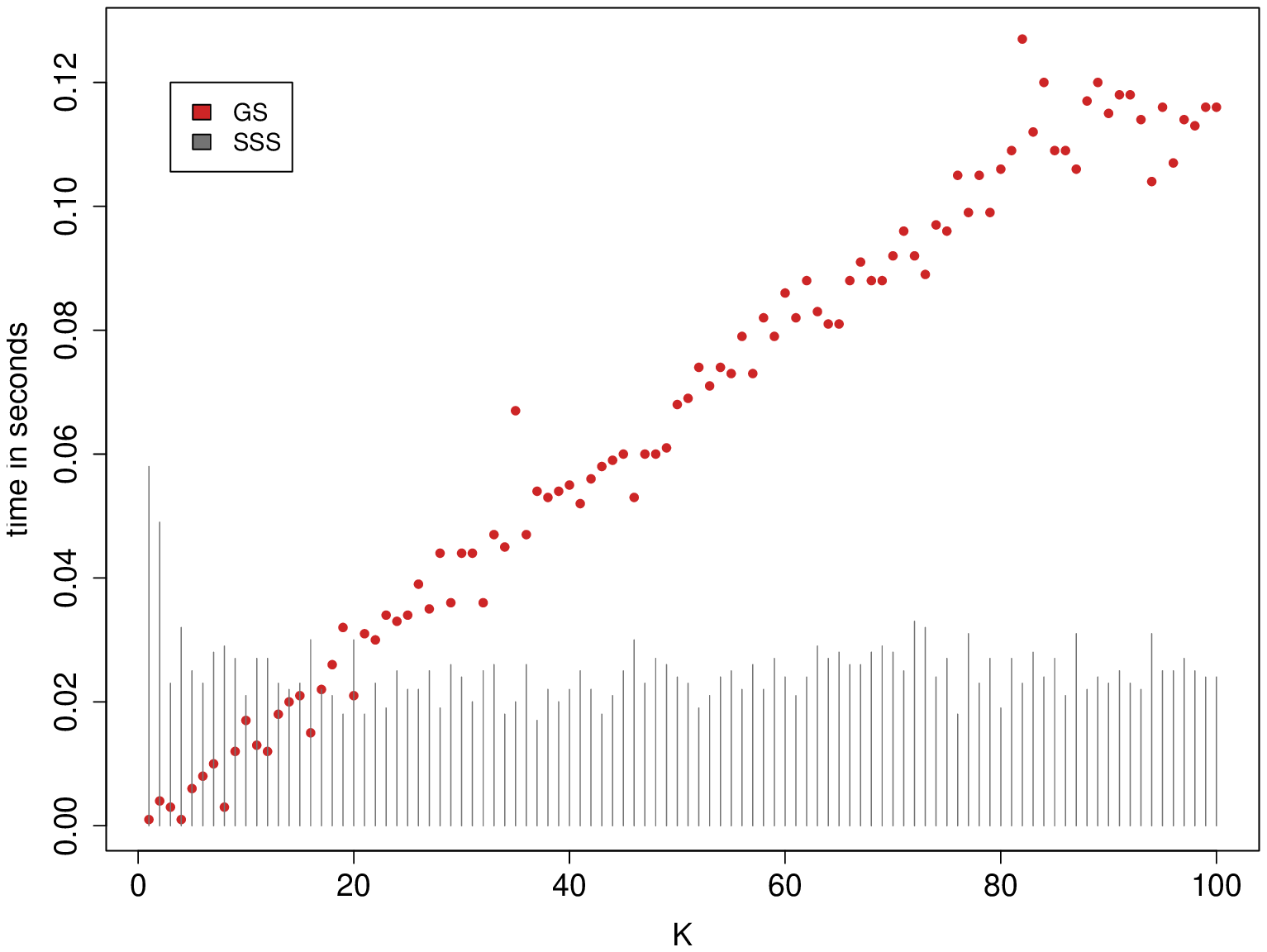}
         \caption{Computing Time}
         \label{fig01d}
     \end{subfigure}
        \caption{Comparing greedy search (GS) with smooth sigmoid surrogate (SSS) in finding the best cutoff point. Panel (a) plots the MSE values (mean squared error) in estimating the true cutoff point $c_0 = 0.5$ on $x \sim \mbox{uniform}[0, 1]$, where SSS with $a \in \{5, 10, 15, \ldots, 100\}$ is compared with GS (the blue dashed line); In panel (b), the density plot of $\hat{c}$ from GS is plotted (in blue), together with those from SSS with various $a$ values. In panel (c), the estimated cutoff points by GS and SSS with $a=50$ are compared with parallel bee swarm plots and boxplots. In panel (d), the computing time is averaged over 20 runs when $x$ is generated from discrete uniform over $\{0, 1, \ldots, K\} /K$ for $K=1, 2, \ldots, 100\}$. }
        \label{fig01-SSS}
\end{figure}

Figure~\ref{fig01-SSS}(a) displays the resulting mean squared error (MSE), defined as $\mathrm{MSE} = \sum_{i=1}^{1000} (\hat{c}_i - 0.5)^2 / 1000$, plotted against different values of $a$. For comparison, the MSE obtained from GS is shown as a blue dotted line. When $a$ is small (e.g., $a = 5$ or $a = 10$), SSS exhibits inferior performance, likely due to the sigmoid function’s limited ability to approximate a threshold function in this range. However, as $a$ increases, the performance of SSS improves markedly, demonstrating both consistency and stability. In particular, for moderate to large values of $a$, SSS consistently achieves a lower MSE than GS.

Figure~\ref{fig01-SSS}(b) presents the density plots of the estimated cutoff values. The orange-shaded region corresponds to the density obtained from GS and serves as a baseline for comparison. The gray curves represent the densities produced by SSS across varying values of the shape parameter $a$. These curves demonstrate consistently stable performance with substantial overlap, particularly as $a$ increases. All density estimates are well-centered around the true cutoff point of 0.5, suggesting that SSS is robust over a broad range of $a$ values. This finding supports the practical strategy of fixing $a$ at a sufficiently large value, particularly when the covariate has been standardized or normalized to the $[0, 1]$ range. In our implementation, we set $a = 50$ as the default value. Figure~\ref{fig01-SSS}(c) presents a bee swarm plot, overlaid with parallel boxplots, comparing the distributions of estimated cutoff points from GS and SSS with $a = 50$. Notably, the GS estimates exhibit greater variability than those from SSS, contributing to the higher mean squared error observed for GS.

In terms of computing time, the number of distinct values of the covariate $Z$, denoted by $K$, emerges as a critical factor influencing runtime. To assess this effect, we simulate $Z$ from a discrete uniform distribution over the set $\{0, 1, \ldots, K\}/K$, varying $K$ from 2 to 100. We then examine how the runtime of GS and SSS scales with increasing $K$. Figure~\ref{fig01-SSS}(d) shows the average computing time (in seconds) across 20 simulation runs for each method. As expected, GS is computationally faster when $K$ is small; however, its runtime increases approximately linearly with $K$. In contrast, SSS demonstrates remarkable stability, maintaining largely unaffected runtime regardless of $K$'s value. Notably, when $K > 20$, SSS consistently outperforms GS in terms of speed, offering substantial computational advantages. Based on these observations, our implementation adopts a hybrid approach: GS is used by default when $K \leq 20$, while SSS is recommended for $K > 20$ to ensure computational efficiency.

\subsection{End-Cut Preference (ECP)}
\label{sec-simulation-ECP}

In regression trees, \citet{Su:2024b} demonstrated that SSS can effectively mitigate the end cut preference (ECP) issue inherent in GS. The ECP problem, originally noted by \citet{Breiman:1984}, refers to the algorithm’s tendency to favor split points near the boundaries of a feature’s range, often resulting in suboptimal or biased splits. We have observed that a similar issue arises when applying GS with logrank test statistics in survival trees.

\begin{figure}[H]
     \centering
     \begin{subfigure}[b]{0.62\textwidth}
         \centering
         \includegraphics[scale=.7, angle=270, width=\textwidth]{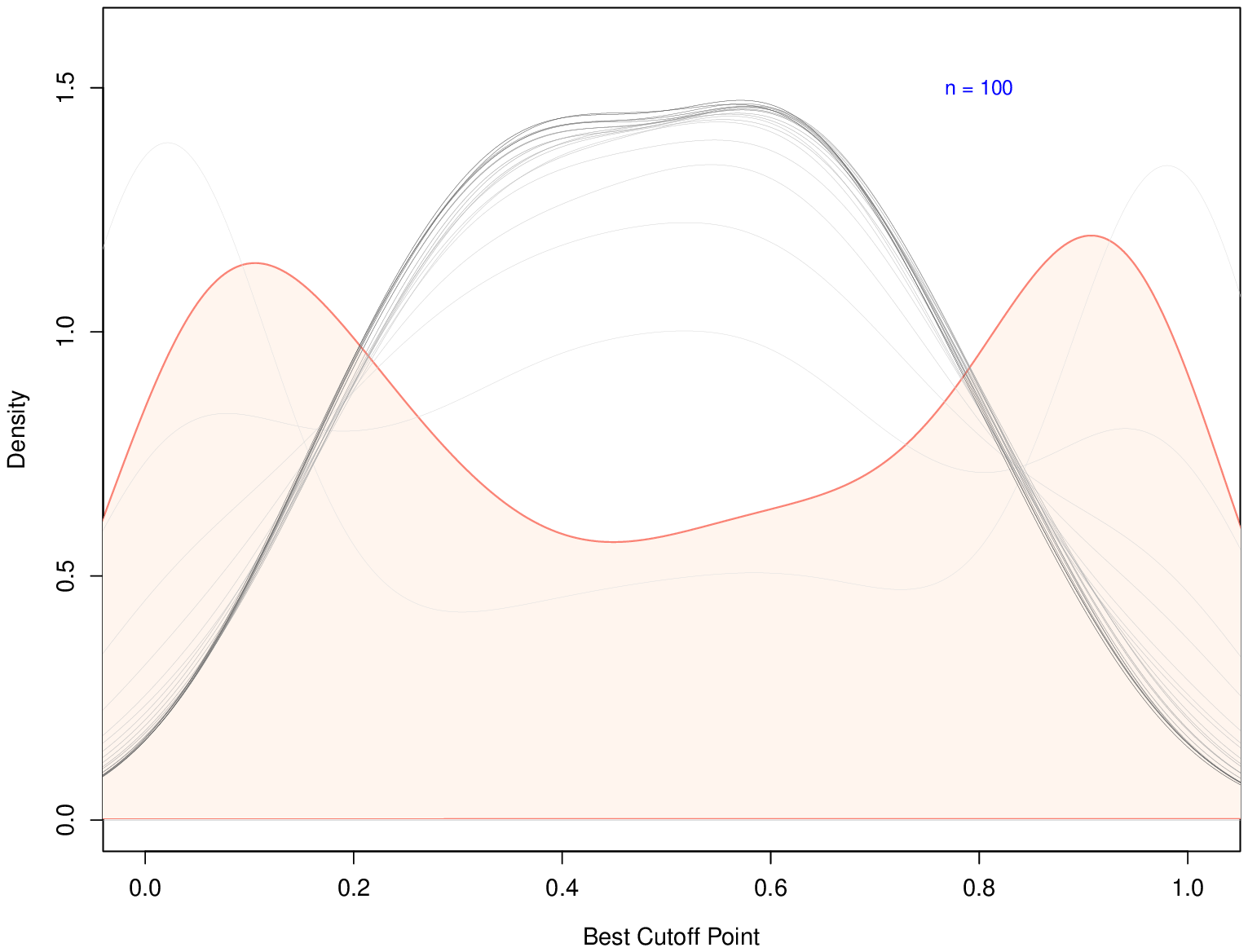}
         \caption{Density}
         \label{fig02a}
     \end{subfigure}
\vspace{0.3in}

     \begin{subfigure}[b]{.62\textwidth}
         \centering
         \includegraphics[scale=.7,angle=270, width=\textwidth]{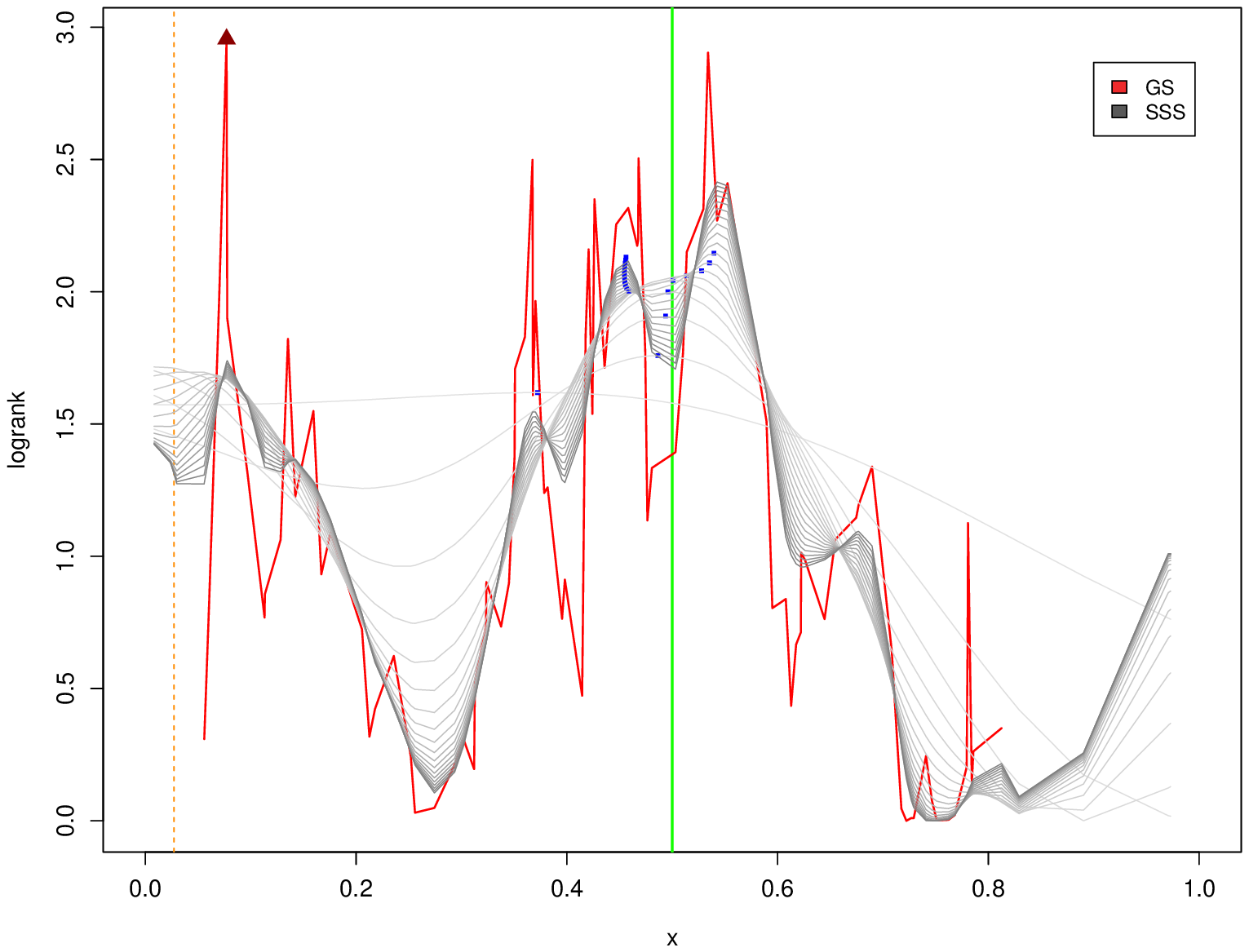}
         \caption{Illustration with One Run}
         \label{fig02b}
     \end{subfigure}
        \caption{Exploring the End-Cut Preference (ECP) issue in greedy search (GS) and smooth sigmoid surrogate (SSS). Panel (a) shows the estimated density of the optimal cutoff points identified by GS (in red) and SSS (in gray) using different $a$ values, $a \in \{5, 10, 15, \ldots, 100\}$. Each density curve is based on 1,000 simulation runs.  Panel (b) illustrates a single simulation run demonstrating how SSS, with a reasonable choice of $a$, addresses the ECP problem observed in GS. The green line represents the true cutoff point $c_0=0.5$; the red triangle marks the best cutoff point identified by the maximum logrank test statistic; the orange dashed line shows the best cutoff point determined by \textbf{rpart}; and the blue dots indicate the best cutoff points identified by the SSS method for various $a$ values. \label{fig02-ECP}}
\end{figure}

To illustrate this, we generate data from Model~(\ref{model-split}) with a weak signal by setting $\beta_1 = -0.1$. The covariate $Z$ follows a uniform distribution on $[0,1]$, and the true cutoff point $c_0 = 0.5$ lies at the center of this range. Figure~\ref{fig02a} displays the density of estimated cutoff points from GS based on 1,000 simulation runs, shown in orange. It is evident that GS frequently misses the true cutoff at the center due to the dominance of ECP, leading to a bimodal distribution with modes near the boundaries. In contrast, the density curves produced by SSS, under various values of the smoothing parameter $a$, are unimodal and concentrated near the center, resembling a mound shape.

Figure~\ref{fig02b} provides further insight by showing the detailed behavior in a single simulation run. The red curve represents the logrank test statistic evaluated at each potential cutoff point and exhibits a highly erratic pattern. Although elevated test statistic values appear near $c_0$ (indicated by a vertical green line), a sudden spike at the lower boundary yields the global maximum. In comparison, SSS acts as a parametric smoothing mechanism that effectively `trims' isolated peaks. Since the spike at the boundary is poorly supported by neighboring points, it undergoes heavier penalization than the smoother peaks near the center. As a result, the maximum shifts back toward the true cutoff point.

In summary, SSS also alleviates the end cut preference issue in survival trees. Further theoretical investigation is warranted to provide rigorous support for this promising empirical observation.

\subsection{Variable Selection Bias}
\label{sec-simulation-VSB}

Next, we evaluate the performance of the proposed intersected validation (IV) method in addressing variable selection bias, and compare it with the GS and GUIDE approaches. The issue of variable selection bias has traditionally been studied under the null setting, where the response is is independent of all predictors, and much of the literature has focused exclusively on this scenario \citep{Loh:2002, Hothorn:2006}. However, variable selection bias also arises in non-null settings. In this study, we investigate both null and non-null cases and show that analysis under non-null settings provides additional insight into the effectiveness of various bias correction methods.

To conduct this evaluation, we generate data from the following model:
\begin{equation}
\label{model-VSB}
\lambda(t) = \exp \left(\beta_0 + \sum_{j=1}^5 \beta_j x_j \right),
\end{equation}
where $\beta_0 = -1$ and the covariates are defined as:
$$ \left\{ \begin{array}{l}
z_1 \sim \text{Bernoulli}(p=0.5)$ \mbox{~with~} $x_1 = z_1; \\
z_2 \sim \text{Discrete Uniform}\{1/10, 2/10, \ldots, 10/10\} \mbox{~with~} x_2 = I(z_2 \leq 0.5); \\
z_3 \sim \text{Discrete Uniform}\{1/50, 2/50, \ldots, 50/50\} \mbox{~with~} x_3 = I(z_3 \leq 0.5); \\
z_4 \sim \text{Uniform}[0, 1] \mbox{~with~} x_4 = I(z_4 \leq 0.5) \\
z_5 \sim \text{Discrete Uniform}\{A, B, \ldots, J\} \mbox{~with~} x_5 = I\left(z_5 \in \{A, B, C, D, E\}\right) \\
\end{array}
\right.$$
By design, the number of unique values increases from $z_1$ through $z_4$, while $z_5$ is a categorical variable with 10 distinct levels. Each $x_j$ represents a binary split derived from the corresponding $z_j$, for $j = 1, \ldots, 5$. The sample size is fixed at $n = 200$, with a censoring rate of 50\%. For each configuration, 1,000 simulation replicates are conducted.

We begin by examining the null scenario by setting $\beta_j = 0$ for $j = 1, \ldots, 5$. Figure~\ref{fig03-Bias}(a) displays bar plots showing the percentage of times each predictor is selected by the different methods. As expected, the bar plots for GS reveal a strong variable selection bias: the probability of selection increases substantially with the number of distinct values a variable possesses. For example, the continuous variable $x_4$ yields up to $(n-1) = 199$ possible cutoff points, while the categorical variable $z_5$ has 10 levels, resulting in $2^{10-1} - 1 = 511$ possible binary splits. Consequently, $z_5$ is selected most frequently under GS. GUIDE, in contrast, treats binary and nominal variables in their native forms, but discretizes each continuous variable into four categories to enable the use of a $\chi^2$ test for variable selection. This strategy balances the selection probabilities among predictors. As shown in Figure~\ref{fig03-Bias}(a), GUIDE performs well in the null setting, with each variable selected approximately 20\% of the time. The proposed intersected validation (IV) method also demonstrates strong performance, as it approximately equalizes the selection probabilities across all predictors.

In the second study, we examine how these methods perform when the binary predictor $z_1$ carries a signal. We vary $\beta_1$ from 0.0 to 1.5 in steps of 0.1, keeping all other slopes at zero, and record the selection frequency of $z_1$. As shown in Figure~\ref{fig03-Bias}(b), GUIDE is the most responsive to weak signals. GS initially struggles due to bias but becomes highly effective once $\beta_1 > 1.0$, selecting $z_1$ over 90\% of the time. The IV approach offers intermediate performance, falling between GUIDE and GS. 

To gain further insight, we design a third study by setting $\beta_1 = \beta_3 = \beta_5 = 1$ and $\beta_2 = \beta_4 = -1$. In this configuration, each predictor contributes a binary split of equal strength, either positively or negatively. As a result, under an unbiased splitting mechanism, all predictors should have an equal probability of being selected. The bar plots in Figure~\ref{fig03-Bias}(c) again show that GS suffers from selection bias, favoring variables with more distinct values. Interestingly, GUIDE exhibits a new form of bias—this time in favor of binary predictors. The strategy that effectively mitigates bias under the null scenario in GUIDE appears to induce a bias toward binary predictors in non-null settings. This explains GUIDE's superior performance in Figure~\ref{fig03-Bias}(b). In contrast, the IV method maintains roughly balanced selection frequencies, successfully mitigating bias in this setting.

The IV approach directly evaluates binary splits, making the selected variable more relevant within the tree structure. Unlike cross-validation, allowing overlap between training and validation data improves its power. Overall, IV proves to be robust and reliable across all examined scenarios, showing great promise as a method for addressing variable selection bias.

\subsection{Tree Modeling}
\label{sec-simulation-tree}
In this subsection, we compare the performance of SurvTreeFuL with the CART approach implemented in the \textbf{rpart} package \citep{Therneau:1990} in R \citep{R:2025}, which is based on martingale residuals.

\renewcommand{\tabcolsep}{6.5pt}
\renewcommand{\arraystretch}{1.1}
\renewcommand{\baselinestretch}{1.0}
\begin{table}[H]
\centering
\caption{Simulation Models Used for Comparing SurvTreeFuL and \texttt{rpart}}
\begin{tabular}{cll}
\hline 
Model & Name & Hazard Function \\
\hline
A & Null & $\lambda_i(t) = \exp(-1)$ \\[4pt]
B & Tree1 & $\lambda_i(t) = \exp\left( -1 + z_{1i} + 2 \cdot I\left\{z_{2i} \leq 0.5 \right\} \right)$ \\[4pt]
C & Tree2 & $\lambda_i(t) = \exp\left( -1 + 3 \cdot z_{1i} \cdot I\left\{0.25 \leq z_{2i} \leq 0.75 \right\} \right)$ \\[4pt]
D & Tree3 & $\lambda_i(t) = \exp\left( -1 + 4 \cdot I\left\{ \sin(6 \pi z_{2i}) \geq 0 \right\} \right)$ \\[4pt]
E & Linear & $\lambda_i(t) = \exp\left( -1 + 3 \cdot z_{2i} - 3 \cdot z_{6i} \right)$ \\[4pt]
F & KAN & $\lambda_i(t) = \exp\left( -1 + 2 \cdot \sin(2\pi z_{2i}^2) + 2 \cdot \sin(2\pi z_{6i}^2) \right)$ \\[4pt]
G & Non-PH & $T_i = \exp\left( -1 + z_{1i} + 2 \cdot I\left\{z_{2i} \leq 0.5 \right\} + \varepsilon_i \right),$ with  $\varepsilon_i \sim \text{logistic}(0, 1)$ \\
\hline
\end{tabular}
\label{tbl-SimModels}
\end{table}
We generate data from seven models, as summarized in Table~\ref{tbl-SimModels}. Each dataset involves seven predictors, denoted by $z_1$ through $z_7$, with the following distributions: $z_1$, $z_4$, and $z_5$ are binary variables independently drawn from a Bernoulli distribution with $p = 0.5$; $z_2$, $z_6$, and $z_7$ are continuous variables independently drawn from a Uniform$[0,1]$ distribution; and $z_3$ is a categorical variable drawn from a discrete uniform distribution over the set $\{A, B, C, D, E\}$.

Model A is a null model and serves to evaluate each method's ability to avoid detecting false signals. Model B represents a standard tree structure with four terminal nodes arranged in a two-level hierarchy, each corresponding to a distinct hazard rate; thus, no node fusion is needed. Model C also has a tree structure, yet with an interaction term. It consists of four leaves at four different depths that naturally form two homogeneous groups, making node fusion necessary. Model D defines a hazard function with an up-down pattern, represented by a tree with six terminal nodes grouped into two sets, again requiring fusion. Model E is a linear exponential model with two continuous predictors. Model F is a nonlinear Cox model following the Kolmogorov–Arnold Networks (KAN) structure \citep{Liu:2024}. 
Finally, since all previous models are based on the Cox proportional hazards (PH) assumption, we introduce Model G as a non-PH alternative. Specifically, Model G adopts a similar tree structure to that of Model B, but it is embedded within a log-logistic accelerated failure time (AFT) model. These models cover a broad spectrum of functional forms and structural complexities, allowing for a comprehensive comparison of SurvTreeFuL and \texttt{rpart}.

We consider a sample size of $n = 600$ with a censoring rate of 50\%. Two strategies are employed to determine the final model: the test-sample approach and $V$-fold cross-validation with $V = 10$. In the test-sample setting, 400 observations are used for training, and the remaining 200 are used for validation. For each model configuration, 200 simulation replicates are conducted.

Three types of performance metrics are used. The first metric is the size of the final model, defined as the number of terminal nodes (leaves) in the CART model or the number of fused groups in the SurvTreeFuL model. We report both the average model size and its standard deviation across the 200 simulation runs. The second metric assesses variable selection performance. We define a model as \emph{inclusive} if all variables used in the splits are truly important, \emph{exclusive} if all truly important variables appear in the model, and \emph{accurate} if it satisfies both the inclusive and exclusive conditions. For each method, we report the frequencies of inclusive, exclusive, and accurate selections. The third metric evaluates predictive performance. To this end, an independent test dataset of size 1,000 is generated from the same underlying model. Each final model is applied to this test set to compute the predicted deviance and concordance.

Table~\ref{tbl02-TestSample} summarizes the results from 200 simulation runs using the test-sample method. Overall, both \texttt{rpart} and SurvTreeFuL provide reasonable approximations to the underlying models. However, SurvTreeFuL consistently yields more parsimonious models than \texttt{rpart}, as reflected in the smaller average model sizes. This is particularly important for Models C and D, where node fusion is necessary to correctly identify the true model structure. Importantly, the parsimony achieved by SurvTreeFuL does not come at the expense of predictive performance. In fact, SurvTreeFuL generally exhibits lower mean deviance values across all Cox proportional hazards (PH) models, with especially notable improvements in Models C and D, precisely the settings where node fusion is essential. Moreover, the standard deviations of deviance under SurvTreeFuL are typically smaller, suggesting more stable performance across simulation runs. The only exception is Model G, the log-logistic accelerated failure time (AFT) model, where \texttt{rpart} slightly outperforms SurvTreeFuL in terms of deviance. This can be attributed to the fact that SurvTreeFuL uses the log-rank test statistic for splitting, which inherently relies on the PH assumption. In contrast, the martingale residual-based approach in \texttt{rpart} is less sensitive to this assumption. It is also worth noting that the concordance measure appears less informative in differentiating between the two methods, with both methods producing very similar values. This may be partly due to the large number of tied pairs in tree-based models, which dominate the denominator in the concordance calculation and diminish its discriminative power as a performance metric. In terms of variable selection, SurvTreeFuL consistently outperforms \texttt{rpart}, particularly in the tree-structured models (Models B, C, and D) and the linear model (Model E). In these settings, SurvTreeFuL achieves substantially higher rates of accurate variable selection. This improvement further underscores the value of model parsimony in enhancing interpretability and selection reliability.

Table~\ref{tbl03-CV} presents the summarized results from 200 simulation runs using the 10-fold cross-validation (CV) method. The overall patterns and conclusions are largely consistent with those obtained from the test sample approach. Our proposed method, SurvTreeFuL, tends to produce more parsimonious models without sacrificing predictive performance. For scenarios such as Models C and D, where leaf fusion is necessary, SurvTreeFuL successfully achieves this goal, as reflected in the smaller model sizes. 

On the other hand, the 10-fold CV results reveal some noteworthy differences compared to those obtained from the test sample method, as shown by comparing Table~\ref{tbl02-TestSample} and Table~\ref{tbl03-CV}. First, the performance metrics under the $V$-fold CV approach generally outperform those from the test sample method, demonstrating lower deviance, higher concordance, and improved accuracy in variable selection. This improvement can be attributed to the larger effective sample size available in the $V$-fold CV framework, which enhances the stability and reliability of the estimates. Second, the relative performance gains of SurvTreeFuL over \texttt{rpart} are slightly diminished in the 10-fold CV setting. This is particularly evident in the average deviance and the percentage of accurately identified variables. Even in Models C and D, where SurvTreeFuL is expected to outperform \texttt{rpart}, the advantages, though still observable, are less pronounced. This attenuation may also be partially due to the increased sample size in CV, which inherently reduces the room for improvement.

\subsection{Bootstrap Bias Correction}
\label{sec-simulation-BBC}

To evaluate the performance of the bootstrap bias correction (BBC) method proposed in Section~\ref{sec-BBC}, we conduct a simulation study using training data $\mathcal{D}$ generated under the Tree2 setting, i.e., Model C in Table~\ref{tbl-SimModels}. The training set consists of $n = 600$ observations. To enhance computational efficiency, the final tree $\mathcal{T}^\star$ and its corresponding grouping structure are determined using an independent validation dataset of size $n = 400$.

Based on the training data $\mathcal{D}$, we fit a Cox proportional hazards model in which the predictor is a categorical variable induced by the groupings from $\mathcal{T}^\star$. The resulting coefficient estimates and their standard deviation (SD) estimates are denoted by $\{\hat{\beta}_k, \hat{\sigma}_k : k = 2, \ldots, K^\star\}$, where $K^\star$ is the number of groups formed by node fusion in $\mathcal{T}^\star$. Next, bias-corrected versions of these estimates are computed using $B = 20$ bootstrap samples. Additionally, an independent test dataset $\mathcal{D}'$ of size $n' = 100{,}000$ is generated from the same model. This test set is passed through the tree $\mathcal{T}^\star$ to recompute the ``true" values of $\hat{\beta}_k$ and $\hat{\sigma}_k$ based on a large-sample approximation.

Figure~\ref{fig04-BBC} summarizes the results across 100 simulation replications. Panel (a) plots the coefficient estimates $\hat{\beta}_k$ from the training data against those computed from the test data, while panel (b) does the same for the SD estimates $\hat{\sigma}_k$. Both panels indicate that the training-based estimates are systematically biased upward. Panels (c) and (d) show the corresponding results after applying the BBC method; the corrected estimates clearly demonstrate substantial reduction in bias.

Recall that Model C is a tree-structured model with two underlying groups. Under perfect model recovery, we would expect to obtain a single non-zero coefficient $\beta = 3$ in each simulation run. As shown in Figure~\ref{fig04-BBC}, most $\hat{\beta}_k$ values are indeed concentrated around 3. Panels (e) and (f) present density contour plots of $\hat{\beta}_k$ and $\hat{\sigma}_k$, respectively, illustrating clustered patterns that likely result from underfitting or overfitting of the tree and group structures in some simulation runs.

We repeat the same experiment using Model E (a linear model) and Model F (a nonlinear model), as defined in Table~\ref{tbl-SimModels}. The corresponding results are provided in the Supplement; see Figure~\ref{figII-BBC}. In both cases, the BBC method performs robustly and effectively mitigates the observed upward bias.

\section{Application}
\label{sec-example}

To further illustrate the utility of our approach, we apply the proposed methods to data from the Alzheimer’s Disease Neuroimaging Initiative (ADNI) study \href{https://adni.loni.usc.edu/}{(https://adni.loni.usc.edu/)}. The ADNI dataset includes magnetic resonance imaging (MRI), positron emission tomography (PET), biological markers, as well as clinical and neuropsychological assessments, which together provide a comprehensive view of Alzheimer’s disease progression. Participants are classified into three diagnostic categories: cognitively normal, mild cognitive impairment (MCI), and Alzheimer’s disease (AD). Following \citet{Li:2017}, we restrict our analysis to individuals diagnosed with MCI at baseline, and define the failure time as the time to progression from MCI to AD. Our research objective is to identify key neuropsychological, functional, behavioral, neuroimaging, clinical, and genetic factors that influence the progression from MCI to AD, thereby enabling the identification of high-risk subgroups who may benefit from enhanced care and targeted research efforts.

Our study includes 991 subjects whose baseline cognitive status was classified as MCI. In the tree-based modeling analysis, we included 19 baseline covariates as potential splitting variables, following the approach in \citet{Yi:2022}. These covariates are: baseline CDR-SOB score (CDRSB\_bl), Alzheimer’s Disease Assessment Scale–Cognitive 13 (ADAS-Cog13) score
(ADAS13\_bl), Rey Auditory Verbal Learning Test immediate
recall (RAVLT\_immediate\_bl), Rey Auditory Verbal Learning Test learning score (RAVLT\_learning\_bl), Rey Auditory Verbal Learning Test percent forgetting (RAVLT\_perc\_forgetting\_bl), Rey Auditory Verbal Learning Test forgetting score (RAVLT\_forgetting \_bl), Mini-Mental State Examination (MMSE) score (MMSE\_bl), Functional Activities Questionnaire score (FAQ\_bl), MRI volumetric data of the ventricles (Ventricles\_bl), hippocampus (Hippocampus\_bl), entorhinal cortex (Entorhinal\_
bl), Intracranial(ICV\_bl), Fusiform gyrus (Fusiform\_bl), Middle temporal gyrus (MidTemp\_bl), Whole brain(WholeBrain\_bl), age (AGE), education (Education), ApoE4 genotype (Apoe4), and gender (Gender).

We compare the performance of the proposed SurvTreeFuL method with that of the standard CART algorithm, implemented in \texttt{rpart}. Figures~\ref{fig05-AZ}(a) and~\ref{fig06-AZ}(b) present the final CART tree (selected by the 1-SE rule) and the final SurvTreeFuL tree, respectively. The latter is derived by applying fused regularization to the CART tree structure. While the CART tree yields six terminal nodes, these are consolidated into three distinct groups in the SurvTreeFuL tree, as indicated by the leaf colors. This highlights SurvTreeFuL’s ability to refine CART, producing a more parsimonious and interpretable model.

Figures~\ref{fig05-AZ}(b) and~\ref{fig06-AZ}(b) present the Kaplan–Meier survival curves for the subgroups identified by the CART and SurvTreeFuL models, respectively. In Figure~\ref{fig05-AZ}(b), the transparent survival curves corresponding to subgroups 5, 6, and 9 follow similar trajectories; these are combined into a single subgroup in Figure~\ref{fig06-AZ}(b), represented by the green survival curve. Likewise, the survival curves for subgroups 14 and 15 in Figure~\ref{fig05-AZ}(b) are merged into a single subgroup in Figure~\ref{fig06-AZ}(b), shown as the blue survival curve. These results underscore the advantages of SurvTreeFuL in producing more parsimonious models with well-separated survival profiles across subgroups.  

For interpretation of the SurvTreeFuL model in Figure~\ref{fig06-AZ}(b), the first split is determined by the baseline ADAS-Cog13 score. The subgroup with a lower baseline ADAS-Cog13 score (ADAS-Cog13 $< 15.84$), larger hippocampal volume (Hippocampus $\geq 6635$), and lower baseline CDR-SOB score (CDRSB\_bl $< 1.25$) is characterized by better global cognitive function, minimal functional impairment, and reduced neurodegeneration. Individuals in this group are less likely to progress from mild cognitive impairment (MCI) to Alzheimer’s disease (AD). In contrast, the subgroup with a higher baseline ADAS-Cog13 score (ADAS-Cog13 $\geq 15.84$) and an elevated Functional Activities Questionnaire score (FAQ\_bl $\geq 0.5$) is identified as high-risk for progression from MCI to AD. These findings align with the results reported by \citet{Ewers:2012} and \citet{Moradi:2015}.  

We also applied the SurvTreeFuL model to the dataset with all proposed features activated. The resulting final tree has seven terminal nodes, which are subsequently merged into three groups, as shown in Figure~\ref{fig07-AZ}(a). The corresponding Kaplan--Meier survival curves for these groups are presented in Figure~\ref{fig07-AZ}(b). In this setting, the final SurvTreeFuL tree begins with a split on the Functional Activities Questionnaire (FAQ). The subgroup characterized by lower FAQ scores (FAQ\_bl $< 2.09$), lower ADAS-Cog13 scores (ADAS-Cog13 $< 18.94$), and larger hippocampal volume (Hippocampus $\geq 6502.82$) is at lower risk of progressing from mild cognitive impairment (MCI) to Alzheimer’s disease (AD).  

Compared with the SurvTreeFuL tree initialized from the CART structure, the log-rank test-based tree introduces the genetic marker ApoE4 as a splitting variable. As noted by \citet{Corder:1993}, ApoE4 is a major genetic risk factor for late-onset AD, and non-carriers are generally regarded as having a lower genetic predisposition to the disease. Within the subgroup of ApoE4 non-carriers, our model identifies a high-risk subgroup defined by smaller midtemporal volume (MidTemp\_bl $< 15{,}921.1$). This finding indicates that even among individuals with lower genetic risk, midtemporal atrophy can help identify populations at elevated risk of conversion from MCI to AD. As shown in Figure~\ref{fig10}, the SurvTreeFuL model based on smoothed log-rank test statistics also yields three subgroups with well-separated Kaplan--Meier survival curves, consistent with the separation observed in the SurvTreeFuL model initialized using CART.  

To summarize and further compare the two grouping structures produced by SurvTreeFuL, we report the estimated hazard ratios and standard deviations for each group, both with and without bootstrap bias correction (BBC). In both cases, the group with the lowest hazard serves as the reference category. The results are displayed in Table~\ref{tbl04-ADNI}, where Tree~I corresponds to the final SurvTreeFuL model initialized from CART, and Tree~II corresponds to the stand-alone SurvTreeFuL model. Bias-corrected estimates are obtained using $B = 25$ bootstrap samples. The results show that both models produce well-separated groups. Moreover, the BBC consistently reduces the upward bias, leading to smaller hazard ratio estimates and larger $p$-values compared with the uncorrected estimates.

\section{Discussion}
\label{sec-discussion}
In this paper, we introduced \textit{SurvTreeFuL}, an enhanced survival tree methodology designed to address persistent challenges in modeling censored survival data. Our approach features three key innovations. First, to improve the efficiency and stability of split selection, we reformulated the traditional greedy search by employing a Smooth Sigmoid Surrogate (SSS) for threshold approximation. This approach not only accelerates the search process but also effectively mitigates the long-standing end-cut preference problem. In addition, we proposed an intersected validation (IV) strategy to reduce variable selection bias, which is particularly beneficial in settings involving continuous or high-cardinality categorical covariates. Second, we developed a new tree modeling framework that incorporates fused regularization. This allows for the merging of non-adjacent terminal nodes based on hazard similarity, resulting in more parsimonious and interpretable trees. Unlike traditional pruning, this regularization-based approach expands the space of candidate models and enables the discovery of higher-level structure in the data. The resulting fused trees are better suited for scientific interpretation and subgroup stratification. Third, we addressed a crucial challenge in tree-based survival analysis by proposing a bootstrap bias correction (BBC) method that enables valid statistical inference for group-specific hazard summaries. This approach accounts for the adaptiveness and instability of recursive partitioning and adjusts both coefficient estimates and standard errors. Extensive simulation studies demonstrate the reliability and robustness of the proposed method across a range of settings.

This work also opens several promising directions for future research. One natural extension is to apply the proposed methodology to more complex survival data, such as those involving interval censoring, discrete event times, time-dependent covariates, multivariate outcomes, or clustered data structures. Adapting SurvTreeFuL to these contexts would substantially increase its applicability to modern biomedical studies. Another important direction is to relax the proportional hazards (PH) assumption that underlies the current framework. Extending the methodology to alternative models such as the accelerated failure time (AFT) model would make the approach more robust in scenarios where the hazard functions are not proportional over time. Furthermore, recent advances in tree-based inference, including bootstrap calibration \citep{Loh:1991b} and selective inference for recursive partitioning \citep{Neufeld:2022} may be integrated into survival trees and compared to our bootstrap bias correction approach. A systematic evaluation of these approaches would shed light on their relative strengths and limitations and guide best practices in statistical inference in the survival tree context. Finally, the methodologies employed in SurvTreeFuL can be integrated with or extended to variants and adaptations of survival trees.  For instance, the SSS and intersected validation approaches may enhance the construction of survival forests \citep{Ishwaran:2008}. The fused regularization and inference techniques could also be adapted for globally optimized survival trees, such as those developed by \cite{Bertsimas:2022}. Furthermore, building on prior work in interaction survival trees \citep[see, e.g.,][]{Su:2008}, a promising direction would be to extend SurvTreeFuL to detect effect modifiers and treatment heterogeneity in both experimental and observational studies.

\newpage

\renewcommand{\tabcolsep}{4.pt}
\renewcommand{\arraystretch}{1.5}
\renewcommand{\baselinestretch}{1.0}
\begin{landscape}
\begin{table}[h]
\caption{Comparison of rpart and SurvTreeFuL using the Test-Sample Method. Results are based on 200 simulation runs for each model. In each run, a training sample of size $n=400$ and a test sample of size $n'=200$ were generated, with a 50\% censoring rate applied.}
\vspace{.2in}
\centering
\begin{tabular}{cllccccccccccc} \hline \hline
	&		&		&	\multicolumn{3}{c}{Model Size}					&&	\multicolumn{3}{c}{Variable Selection}					&&	\multicolumn{2}{c}{Deviance}			&	Averaged	\\	\cline{4-6} \cline{8-10} \cline{12-13}
\multicolumn{2}{l}{Model}			&	Method	&	Expected	&	Mean	&	SD	&&	Inclusive	&	Exclusive	&	Accurate	&&	Mean	&	SD	&	Concordance	\\	\hline
A	&	Null	&	rpart	&	1	&	1.29	&	1.387	&&		---	& ---			&	0.890	&&	6126.612	&	62.298	&	0.500	\\	
	&		&	SurvTreeFuL	&	1	&	1.12	&	0.461	&&		---	& ---			&	0.925	&&	6119.001	&	5.964	&	0.500	\\	\hline
B	&	Tree1	&	rpart	&	4	&	4.06	&	2.195	&&	0.795	&	0.800	&	0.595	&&	5854.382	&	137.036	&	0.718	\\	
	&		&	SurvTreeFuL	&	4	&	3.81	&	0.582	&&	0.905	&	0.980	&	0.885	&&	5814.080	&	65.573	&	0.724	\\	\hline
C	&	Tree2	&	rpart	&	4	&	5.18	&	3.864	&&	0.730	&	0.875	&	0.605	&&	5875.643	&	182.366	&	0.659	\\	
	&		&	SurvTreeFuL	&	2	&	2.16	&	0.568	&&	0.905	&	0.900	&	0.815	&&	5847.525	&	117.744	&	0.659	\\	\hline
D	&	Tree3	&	rpart	&	6	&	6.12	&	1.995	&&	0.725	&	1.000	&	0.725	&&	5733.084	&	265.376	&	0.716	\\	
	&		&	SurvTreeFuL	&	2	&	2.27	&	0.669	&&	0.940	&	1.000	&	0.940	&&	5652.865	&	130.013	&	0.715	\\	\hline
E	&	Linear	&	rpart	&	Many	&	6.78	&	5.829	&&	0.780	&	0.870	&	0.650	&&	5976.964	&	152.163	&	0.687	\\	
	&		&	SurvTreeFuL	&	Many	&	4.43	&	1.242	&&	0.980	&	0.990	&	0.970	&&	5887.748	&	51.537	&	0.699	\\	\hline
F	&	KAN	&	rpart	&	Many	&	5.64	&	4.236	&&	0.905	&	0.990	&	0.895	&&	5905.910	&	153.317	&	0.718	\\	
	&		&	SurvTreeFuL	&	Many	&	4.57	&	1.197	&&	0.980	&	1.000	&	0.980	&&	5823.924	&	54.114	&	0.733	\\	\hline
G	&	Non-PH	&	rpart	&	4	&	3.33	&	2.091	&&	0.845	&	0.595	&	0.450	&&	5785.006	&	79.479	&	0.692	\\	
	&		&	SurvTreeFuL	&	4	&	3.14	&	0.874	&&	0.870	&	0.720	&	0.600	&&	5807.120	&	95.111	&	0.657	\\	\hline
\end{tabular}
\label{tbl02-TestSample}
\end{table}
\end{landscape}

\renewcommand{\tabcolsep}{4.pt}
\renewcommand{\arraystretch}{1.5}
\renewcommand{\baselinestretch}{1.0}
\begin{landscape}
\begin{table}[h]
\caption{Comparison of rpart and SurvTreeFuL using the $10$-Fold CV Method. Results are based on 200 simulation runs for each model. In each run, a training sample of size $n=600$ was generated, with a 50\% censoring rate applied.}
\vspace{.2in}
\centering
\begin{tabular}{cllccccccccccc} \hline \hline
	&		&		&	\multicolumn{3}{c}{Model Size}					&&	\multicolumn{3}{c}{Variable Selection}					&&	\multicolumn{2}{c}{Deviance}			&	Averaged	\\	\cline{4-6} \cline{8-10} \cline{12-13}
Model	&	Name	&	Method	&	Expected	&	Mean	&	SD	&&	Inclusive	&	Exclusive	&	Accurate	&&	Mean	&	SD	&	Concordance	\\	\hline
A	&	Null	&	rpart	&	1	&	1.06	&	0.320	&&	---	& ---		&	0.965	&&	6122.702	&	57.431	&	0.500	\\	
	&		&	SurvTreeFuL	&	1	&	1.03	&	0.157	&&		---	& ---		&	0.975	&&	6118.424	&	4.152	&	0.500	\\	\hline
B	&	Tree1	&	rpart	&	4	&	4.06	&	1.304	&&	0.830	&	0.960	&	0.790	&&	5814.771	&	89.931	&	0.728	\\	
	&		&	SurvTreeFuL	&	4	&	3.75	&	0.742	&&	0.835	&	0.960	&	0.795	&&	5811.880	&	80.877	&	0.727	\\	\hline
C	&	Tree2	&	rpart	&	4	&	4.29	&	0.773	&&	0.930	&	0.995	&	0.930	&&	5780.929	&	51.679	&	0.684	\\	
	&		&	SurvTreeFuL	&	2	&	2.14	&	0.445	&&	0.950	&	0.995	&	0.945	&&	5776.724	&	51.617	&	0.683	\\	\hline
D	&	Tree3	&	rpart	&	6	&	6.75	&	1.211	&&	0.820	&	1.000	&	0.820	&&	5620.045	&	95.751	&	0.739	\\	
	&		&	SurvTreeFuL	&	2	&	2.40	&	0.935	&&	0.885	&	0.995	&	0.880	&&	5613.812	&	104.488	&	0.736	\\	\hline
E	&	Linear	&	rpart	&	Many	&	6.44	&	2.296	&&	0.810	&	1.000	&	0.810	&&	5901.959	&	107.245	&	0.707	\\	
	&		&	SurvTreeFuL	&	Many	&	4.55	&	1.194	&&	0.810	&	1.000	&	0.810	&&	5900.721	&	102.748	&	0.703	\\	\hline
F	&	KAN	&	rpart	&	Many	&	8.10	&	3.359	&&	0.830	&	1.000	&	0.830	&&	5812.638	&	100.041	&	0.751	\\	
	&		&	SurvTreeFuL	&	Many	&	5.37	&	1.394	&&	0.835	&	1.000	&	0.835	&&	5809.811	&	85.669	&	0.746	\\	\hline
G	&	Non-PH	&	rpart	&	4	&	3.20	&	1.199	&&	0.880	&	0.685	&	0.570	&&	5764.646	&	41.291	&	0.699	\\	
	&		&	SurvTreeFuL	&	4	&	3.03	&	0.856	&&	0.880	&	0.685	&	0.570	&&	5765.030	&	42.005	&	0.698	\\	\hline
\end{tabular}
\label{tbl03-CV}
\end{table}
\end{landscape}

\renewcommand{\tabcolsep}{3.5pt}
\renewcommand{\arraystretch}{1.8}
\renewcommand{\baselinestretch}{1.2}
\begin{table}[h]
\caption{Group summaries from the SurvTreeFuL modeling analysis of the ADNI dataset. Tree~I refers to the SurvTreeFuL model in Figure~\ref{fig06-AZ}, which is built upon the final \texttt{rpart} tree, whereas Tree~II corresponds to the SurvTreeFuL model shown in Figure~\ref{fig07-AZ}.}
\vspace{.2in}
\centering
\begin{tabular}{ccrrrrrcccccc} \hline \hline
& & & & & &  & \multicolumn{6}{c}{Bootstrap Bias Correction (BBC)} \\ \cline{9-13}
Tree & Group & $\beta$ & HR & SE & $Z$ & $p$-value  && $\beta$ & HR & SE & $Z$ & $p$-value \\ \hline
I & 1 & 0.000 & 1.000 & NA      & NA      & NA               && 0.000       & 1.000 & NA        & NA       & NA \\
 & 2 & 2.542 & 12.704 & 0.371  & 6.846  & $7.57 \times 10^{-12}$  && 2.342   & 10.402  & 0.639    & 3.667   & $2.45 \times 10^{-4}$ \\
& 3 & 4.347 & 77.729 & 0.362  & 12.004 & $3.38 \times 10^{-33}$  && 3.960  & 52.441  & 0.612    & 6.470   & $9.78 \times 10^{-11}$ \\
\hline
II & 1 & 0.000 & 1.000 & NA      & NA      & NA              && 0.000 & 1.000 & NA        & NA       & NA \\
& 2 & 2.467 & 11.792 & 0.221  & 11.180 & $5.13 \times 10^{-29}$  && 2.233  & 9.330   & 0.343    & 6.518   & $7.15 \times 10^{-11}$ \\
& 3 & 4.122 & 61.691 & 0.233  & 17.716 & $3.14 \times 10^{-70}$ && 3.780 & 43.797  & 0.548    & 6.897   & $5.32 \times 10^{-12}$ \\
\hline
\end{tabular}
\label{tbl04-ADNI}
\end{table}

\begin{figure}[H]
\renewcommand\arraystretch{1}%
\cellspacetoplimit3pt
\cellspacebottomlimit4pt
\centering
\begin{tabular}[t]{c}
\includegraphics[height=12cm, width=6cm, angle=270]{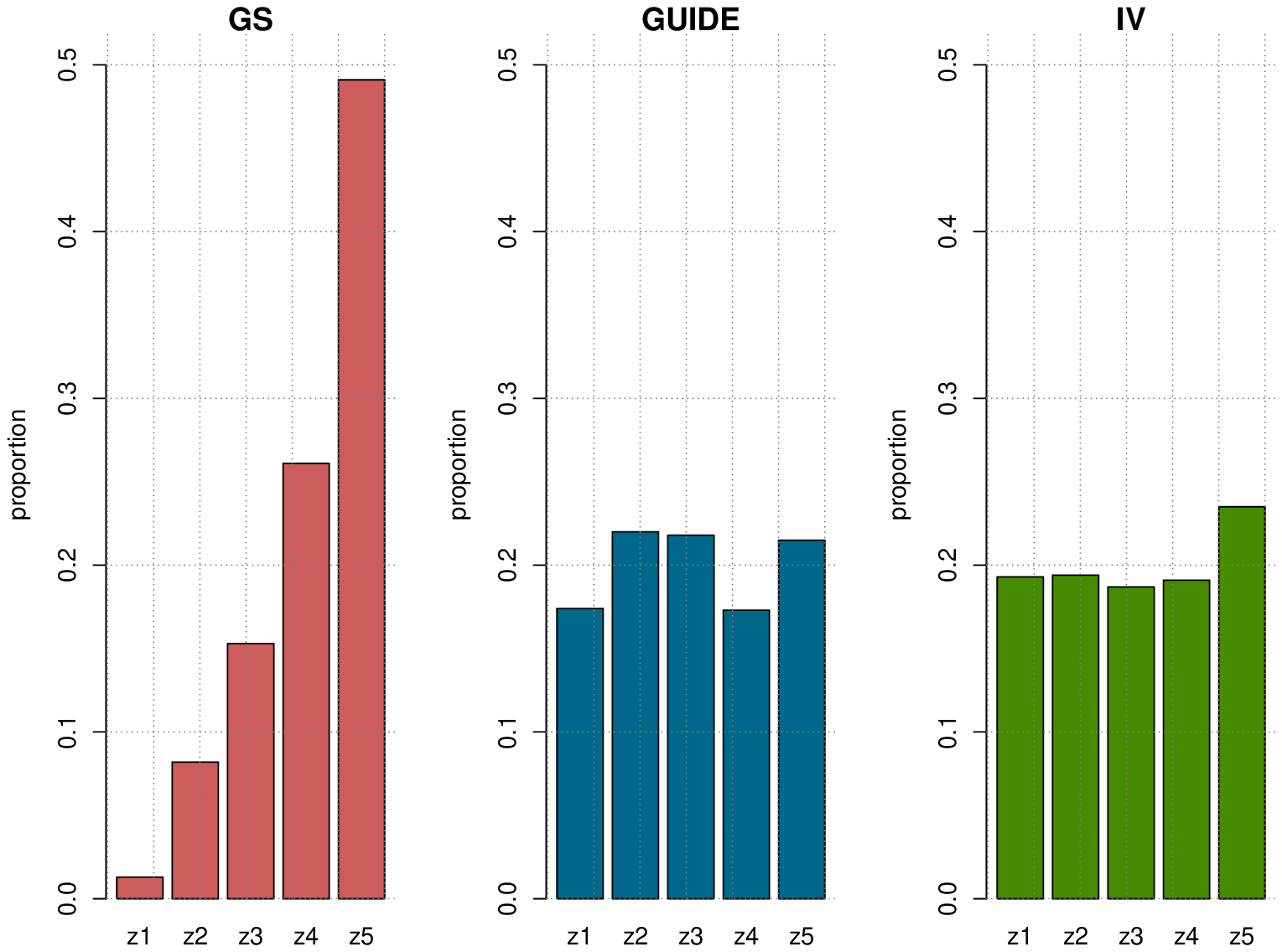} \\
  (a) Null  \\
\includegraphics[height=9cm, width=5.5cm, angle=270]{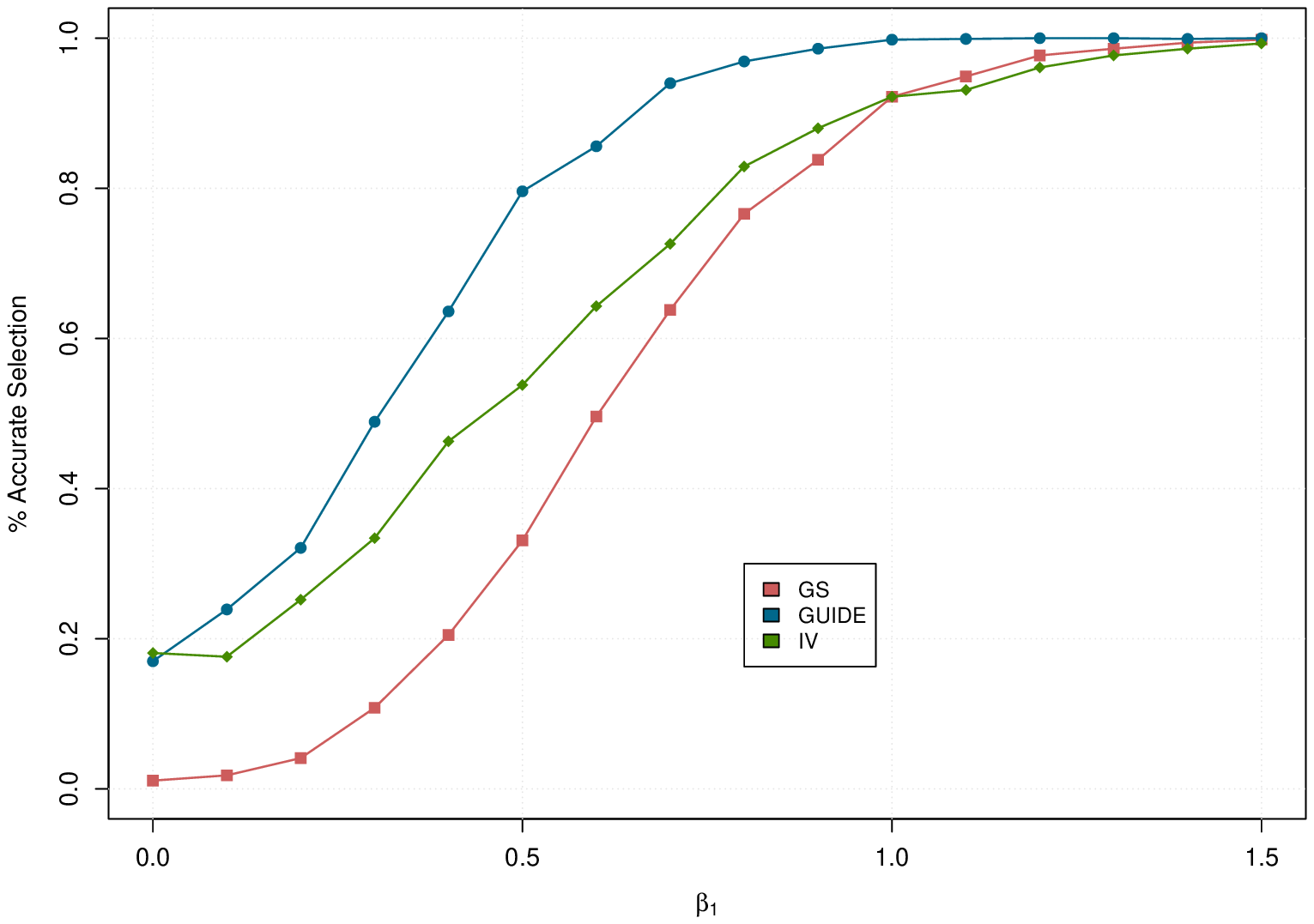} \\
  (b) Signal with Binary Variable  \vspace{0.1in} \\ 
\includegraphics[height=12cm, width=6cm, angle=270]{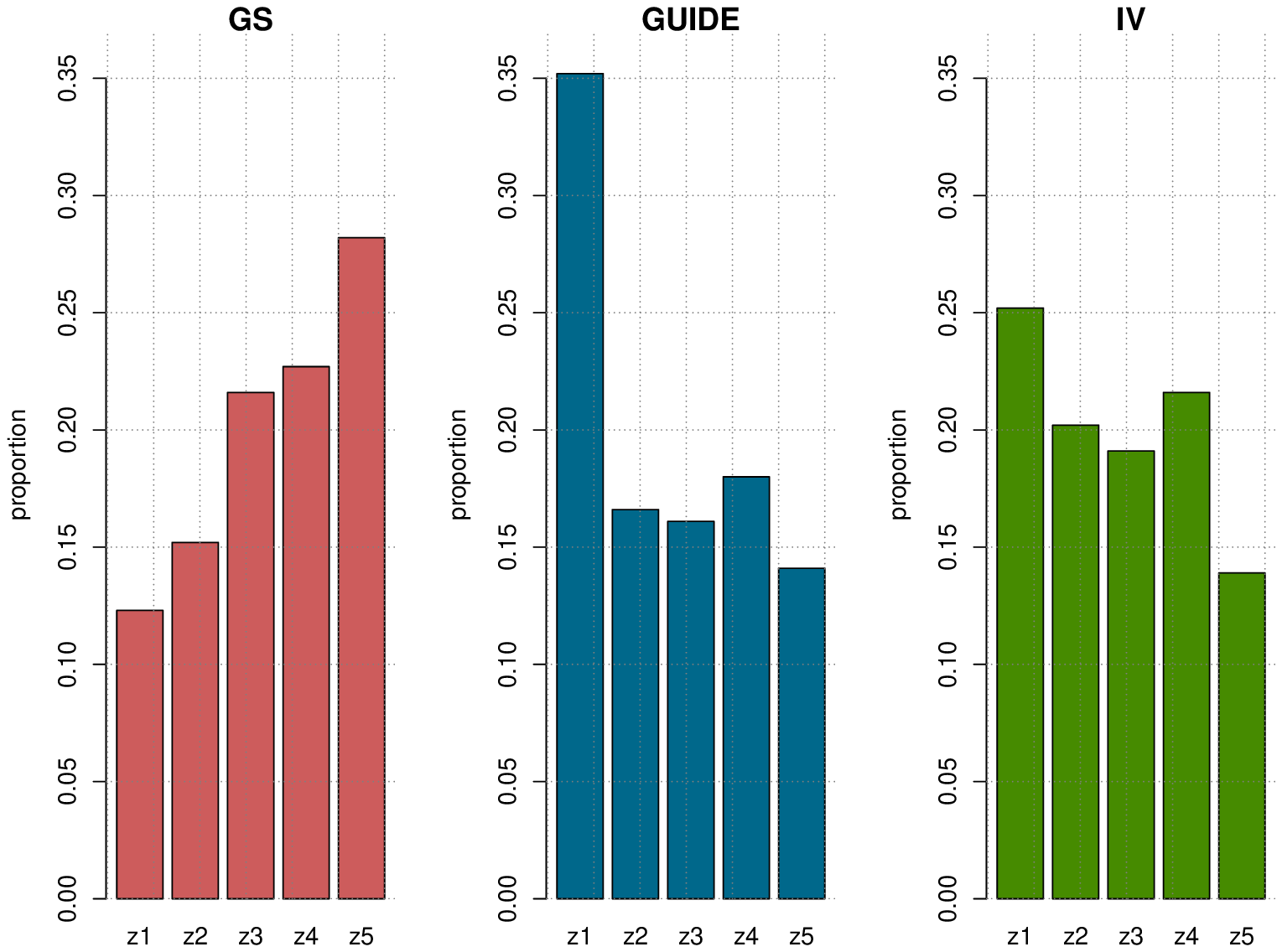}\\
 (c) Non-Null \\
 \end{tabular}
 \caption{Exploring Variable Selection Bias with Three Methods, Greedy Search (GS), GUIDE, and Intersected Validation (IV).   \label{fig03-Bias}}
\end{figure}

\begin{figure}[H]
     \centering
\includegraphics[scale=0.7, angle=0]{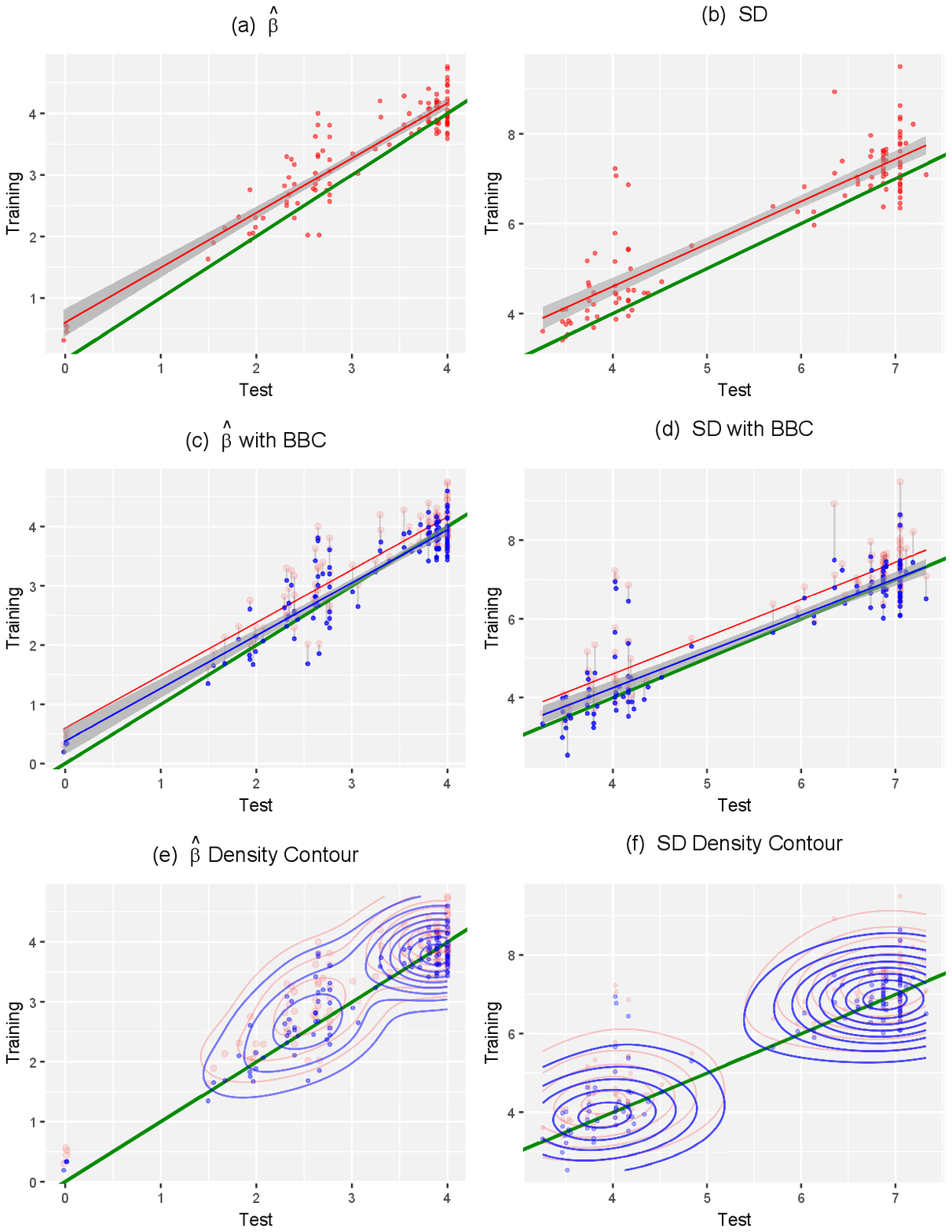}
 \caption{Cox model-based group summaries with bootstrap bias correction (BBC) using data generated from the Tree2 model (Model C in Table~\ref{tbl-SimModels}). Left panels (a, c, e) show training-based estimates ({\color{red}red}), BBC-adjusted estimates ({\color{blue} blue}), and density contour plots for coefficients $\beta$ versus estimates from an independent test sample. Right panels (b, d, f) show the corresponding results for standard deviations (SD). The {\color{green} green} line denotes the reference line $y = x$.
\label{fig04-BBC}}
\end{figure}

\begin{figure}[H]
     \centering
     \begin{subfigure}[b]{1\textwidth}
         \centering
         \includegraphics[scale=0.3, angle=0, width=0.5\textwidth]{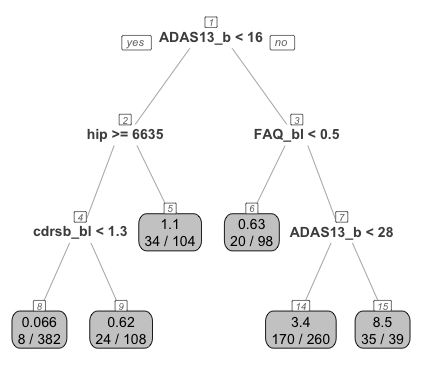}
         \caption{}
         \label{fig05a}
     \end{subfigure}
     
\vspace{0.3in}
     \begin{subfigure}[b]{1\textwidth}
         \centering
         \includegraphics[scale=0.5,angle=0, width=0.85\textwidth]{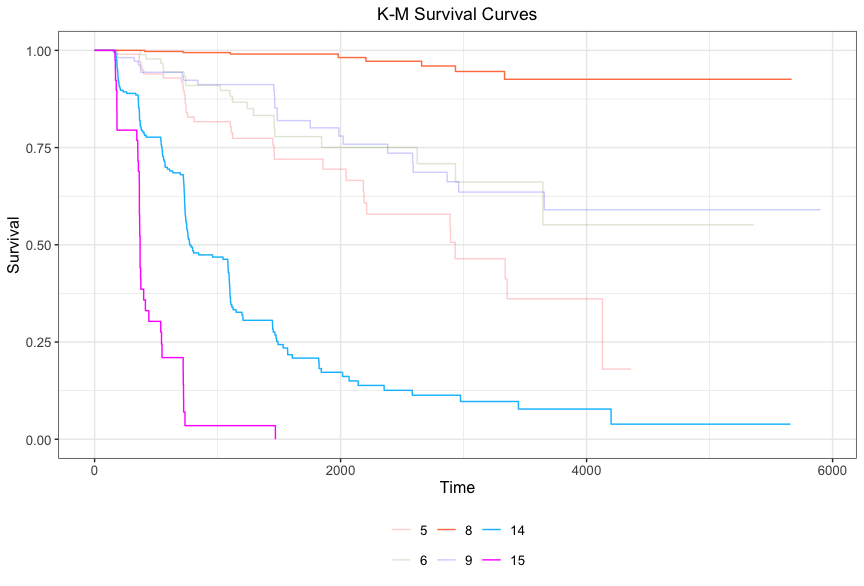}
         \caption{}
         \label{fig05b}
     \end{subfigure}
     \caption{Results of the \texttt{rpart} analysis on the ADNI dataset: (a) the final \texttt{rpart} tree structure; (b) the corresponding Kaplan--Meier (KM) survival curves.}
        \label{fig05-AZ}
\end{figure}

\begin{figure}[H]
     \centering
     \begin{subfigure}[b]{1\textwidth}
         \centering
         \includegraphics[scale=0.35, angle=0, width=0.7\textwidth]{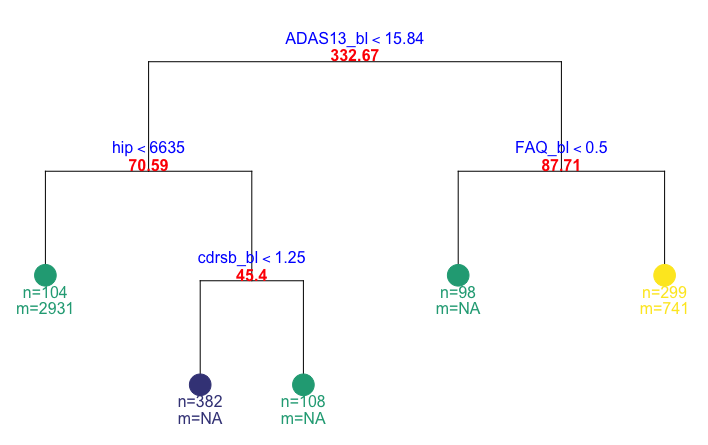}
         \caption{}
         \label{fig06a}
     \end{subfigure}
     
\vspace{0.2in}

     \begin{subfigure}[b]{1\textwidth}
         \centering
         \includegraphics[angle=0, width=0.85\textwidth]{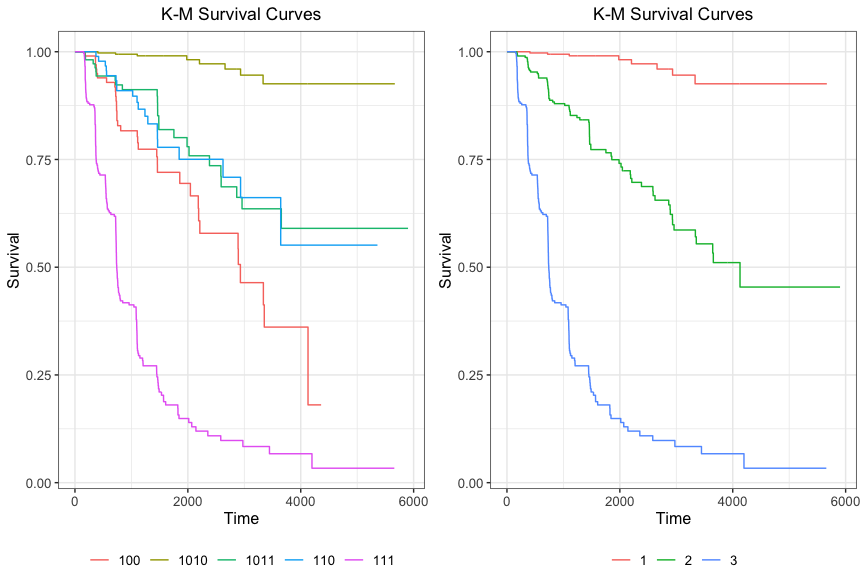}
         \caption{}
         \label{fig06b}
     \end{subfigure}
        \caption{Results of the SurvTreeFuL analysis on the ADNI dataset, building upon the final \texttt{rpart} tree. Panel (a) displays the final SurvTreeFuL tree structure, while Panel (b) presents the Kaplan--Meier (KM) survival curves before and after leaf fusion.}
        \label{fig06-AZ}
\end{figure}

\begin{figure}[H]
     \centering
     \begin{subfigure}[b]{1\textwidth}
         \centering
         \includegraphics[scale=0.45, angle=0, width=0.75\textwidth]{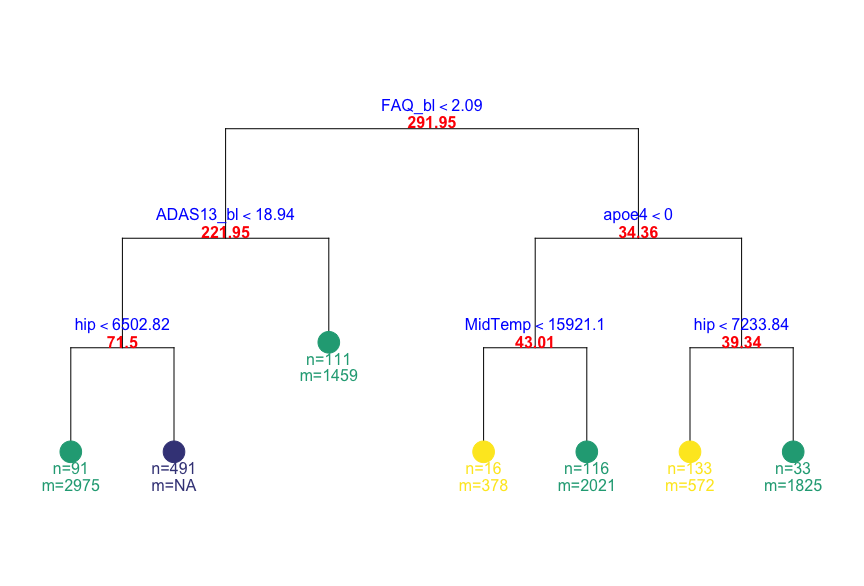}
         \caption{}
         \label{fig07a}
     \end{subfigure}
     
\vspace{0.2in}

     \begin{subfigure}[b]{1\textwidth}
         \centering
         \includegraphics[angle=0, width=0.85\textwidth]{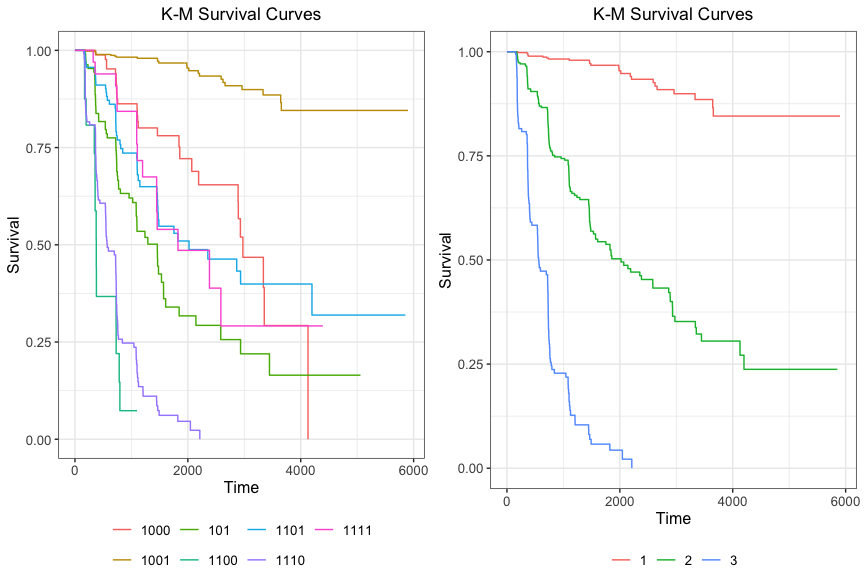}
         \caption{}
         \label{fig07b}
     \end{subfigure}
\caption{Results of the SurvTreeFuL analysis on the ADNI dataset with all features fully activated. Panel (a) shows the final SurvTreeFuL tree structure, and Panel (b) presents the Kaplan--Meier (KM) survival curves before and after leaf fusion.}
        \label{fig07-AZ}
\end{figure}


\newpage
\clearpage \setcounter{page}{1} \setcounter{table}{0}
\setcounter{figure}{0}
\renewcommand{\thetable}{\Roman{table}}
\renewcommand{\thefigure}{\Roman{figure}}
\numberwithin{equation}{section}

\begin{center}
{\Large \textbf{Supplementary Material to}} \\
\vspace{0.1in}
{\Large \textbf{``Enhanced Survival Trees"}}
\end{center}

\appendix
\section{Algorithms}
\label{sec-Alg} 

This section presents several pseudocode algorithms that form key components of SurvTreeFuL. Algorithm~\ref{Algorithm-1} describes the \emph{intersected validation} (IV) approach, which addresses variable selection bias in greedy search procedures. Algorithm~\ref{Algorithm-2} outlines the $V$-fold cross-validation (CV) technique used for tuning parameter selection in the fused regularization step of SurvTreeFuL. Finally, Algorithm~\ref{Algorithm-3} details the bootstrap bias correction procedure for estimating the logarithms of hazard ratios when summarizing the final groups.

\vspace{.2in}
\begin{algorithm}[H]
\caption{Node Partition with Unbiased Variable Selection via Intersected Validation (IV)} \label{Algorithm-1}
\algrule \KwData{Data $\mathcal{D}=\{(T_i, \Delta_i, \bm{z}_i)\}_{i=1}^n$} 
\KwResult{Best binary split $s^\star$} 
\textbf{initialize} $\Delta$ as the stratification factor.\;
\Begin{
	Randomly partition $\mathcal{D} = \bigcap_{k=1}^3 \mathcal{D}_k$ of equal sizes\;
	Let $n_k = |  \mathcal{D}_k|$ with $|\cdot|$ being cardinality\; 
	\Begin{
	Construct $\mathcal{D}^{\mbox{\scriptsize(B)}}_{12}$ of size $n_2+n_3$ with stratified sampling from $\mathcal{D}_1+\mathcal{D}_2$\;
	Form training data $\mathcal{D}'_1 = \mathcal{D}_1 \cup \mathcal{D}^{\mbox{\scriptsize(B)}}_{12}$.
	}
	Using $\mathcal{D}'_1$, determine the optimal cutoff point for each predictor: $s_j = (Z_j, c^\star_j)$\;
	\Begin{ 
	Obtain the out-of-bag (OOB) data $\mathcal{D}^{\mbox{\scriptsize (OOB)}}_2$ from $\mathcal{D}_2$ that were not included in $\mathcal{D}^{(B)}_{12}$\; 
	Let $n'_2 = \left| \mathcal{D}^{\mbox{\scriptsize (OOB)}}_2 \right|$ \; 
	Construct $\mathcal{D}^{\mbox{\scriptsize(B)}}_{23}$ of size $(n-n'_2-n_3)$ with stratified sampling from $ \mathcal{D}_2 + \mathcal{D}_3$\;
	Form validation data $\mathcal{D}'_2  = \mathcal{D}_3 \cup \mathcal{D}^{\mbox{\scriptsize(OOB)}}_2 \cup \mathcal{D}^{\mbox{\scriptsize(B)}}_{23}$\;
}
 	Using $\mathcal{D}'_2$ to validate $s_j$, recompute logrank test statistic at $s_j$ as $Q'(s_j)= Q(c^\star_j)$ for each $Z_j$\;
 \Begin{ 
 	Determine the best splitting variable $j^\star = \argmax_j Q'(s_j)$\;
 	Determine the best split $s^\star = (Z_{j^\star}, c^\star_{j^\star})$ by recomputing $c^\star_{j^\star}$ using $\mathcal{D}$\; 
 	Output the associated logrank test statistic $Q(s^\star)$\;
 	}
 	Partition $h$ into $h_L$ and $h_R$ by $s^\star$.
} \algrule
\end{algorithm}

\clearpage
\IncMargin{1em}
\begin{algorithm}[H]
\caption{$V$-Fold CV for Tuning Parameter Selection.} \label{Algorithm-2}
\SetKwData{Left}{left}\SetKwData{This}{this}\SetKwData{Up}{up}
\SetKwInOut{Input}{input}\SetKwInOut{Output}{output}
\algrule
\KwData{Survival data $\mathcal{D}=\{(T_i, \Delta_i, \bm{z}_i)\}_{i=1}^n$}
\KwResult{Optimal tuning parameter $\lambda^\star$}
\textbf{Initialize} $V$, the number of folds\;
\Begin{
\Begin{
	Grow a large initial tree $\T_0$ with data $\mathcal{D}$\;
	Sort the terminal nodes of $\T_0$ by hazard rate and define dummy vector $\bm{x}_i$ \;
	Apply fused regularization to obtain a regularization path
$ \left\{ (\lambda_m, \widetilde{\bm{\beta}}_m) : m = 1, \ldots, M \right\}$\;
}
	Partition $\mathcal{D} = \bigcup_{v=1}^V \mathcal{D}_v$ and obtain $\mathcal{D}^{(v)} = \mathcal{D} \setminus \mathcal{D}_v$\;
	Initialize $\mbox{D}_m =0$ for $m=1, \ldots , M$\;
    \For{$v \leftarrow 1$ to $V$}{
    \Begin{
        Using $\mathcal{D}^{(v)}$, grow an initial tree $\mathcal{T}_v$  and sort terminal nodes\;
		Apply fused regularization to $\mathcal{D}^{(v)}$ with the same set of tuning parameters $\left\{ \lambda_m: m=1, \ldots, M \right\}$\;
		Obtain solution path $\left\{ \left(\lambda_m, \widetilde{\bm{\beta}}_m^{(v)}, \widehat{\bm{\beta}}_m^{(v)}, \hat{\Lambda}_{0m}^{(v)}(\cdot) \right) : m = 1, \ldots, M \right\}
$\; }

		Send $\mathcal{D}_v$ down to $\mathcal{T}_v$\;
		\For{$m \leftarrow 1$ to $M$}{
				Compute validation deviance $$D_{mv} = 2 \sum_{i \in \mathcal{D}_v} \left[ \hat{\Lambda}_{0m}^{(v)}(T_i) \exp \left( \bm{x}_i^T \widehat{\bm{\beta}}_m^{(v)} \right) 
- \Delta_i \left\{ 1 + \bm{x}_i^T \widehat{\bm{\beta}}_m^{(v)} + \log \hat{\Lambda}_{0m}^{(v)}(T_i) \right\} \right]
\; $$ 
				Update $D_m :=  D_m + D_{mv}$\;
				}
   }
    Determine optimal $\lambda^\star = \lambda_{m^\star},$ with $m^\star = \arg\min_m D_m.$ 
} \algrule
\end{algorithm}

\clearpage
\vspace{.2in}
\begin{algorithm}[H]
\caption{Bootstrap Bias Correction (BBC) for $\beta_k$ Estimate in Group Summary.} \label{Algorithm-3}
\algrule \KwData{Data $\mathcal{D}$ and tree $\T^\star$ with $K^\star$ groups} 
\KwResult{Bias-corrected estimate $\hat{\beta}_k$} 
\textbf{initialize} $B$, the number of bootstrap samples.\;
\Begin{
	 Send $\mathcal{D}$ down to $\T^\star$ and compute \; 
	\Begin{ $\hat{\beta}_k = \hat{\beta}_k(\mathcal{D}, \T^\star)$ for $k=1, \ldots, K^\star$\;
	Group membership vector $\bm{g}$\;
	}
	Initialize $\tau_k = 0$ for $k=1, \ldots, K^\star$\;
Draw $B$ bootstrap samples \;
\For{$b \leftarrow 1$ to $B$}{
    Construct $\T_b$ with $K_b \approx K^\star$ groups, based on $\mathcal{D}_b$ \;
    Send $\mathcal{D}_b$ down to $\T_b$ and compute $\hat{\beta}_{k'}(\mathcal{D}_b, \T_b)$ for $k'=1, \ldots, K_b$ \;
    Send $\mathcal{D}$ down to $\T_b$ and compute: \;
    \Begin{ 
    $\hat{\beta}_{k'}(\mathcal{D}, \T_b)$ for $k'=1, \ldots, K_b$\;
    Group membership vector $\bm{g}_b$ \;
    }
    Compute bias $\tau_{bk'} = \hat{\beta}_{k'}(\mathcal{D}_b, \T_b) - \hat{\beta}_{k'}(\mathcal{D}, \T_b)$ \;
    Form two-way $K^\star \times K_b$ contingency table $\{n_{kk'}\}$ on the basis of $\bm{g}$ and $\bm{g}_b$ \;
    Compute row proportions $p_{kk'} = n_{kk'} / \sum_{k'=1}^{K_b} n_{kk'}$ \;
    \For{$k \leftarrow 1$ to $K^\star$}{
    	update $\tau_k := \tau_k + \sum_{k'=1}^{K_b} p_{kk'} \tau_{kk'}$ \;
    }    
  } 
  \For{$k \leftarrow 1$ to $K^\star$}{
   Average bias $\tau_k := \tau_k/B$\;
   Bias correction $\hat{\beta}_k := \hat{\beta}_k + \tau_k$\;
	}
	Output $\{ \hat{\beta}_k: k=1, \ldots, K^\star \}.$ 
}
\algrule
\end{algorithm}
\vspace{.2in}

\section{Illustration of Coloring and Shearing}
\label{sec-Illustration} 

Figure~\ref{figI-NodeFusion} demonstrates the \emph{coloring} and \emph{shearing} steps of the SurvTreeFuL procedure. A dataset of size $n = 600$ was simulated from Model C in Equation~(3), which has an underlying tree structure with an interaction between a binary variable $Z_1$ and the indicator variable $I(0.25 \leq Z_2 \leq 0.75)$, where $Z_2 \sim \mathrm{Uniform}[0, 1]$. Ideally, the true model yields a final tree with four terminal nodes that form two distinct risk groups.

\afterpage{
\begin{figure}[H]
     \centering
     \begin{subfigure}[b]{0.65\textwidth}
         \centering
         \includegraphics[scale=.7, angle=270, width=\textwidth]{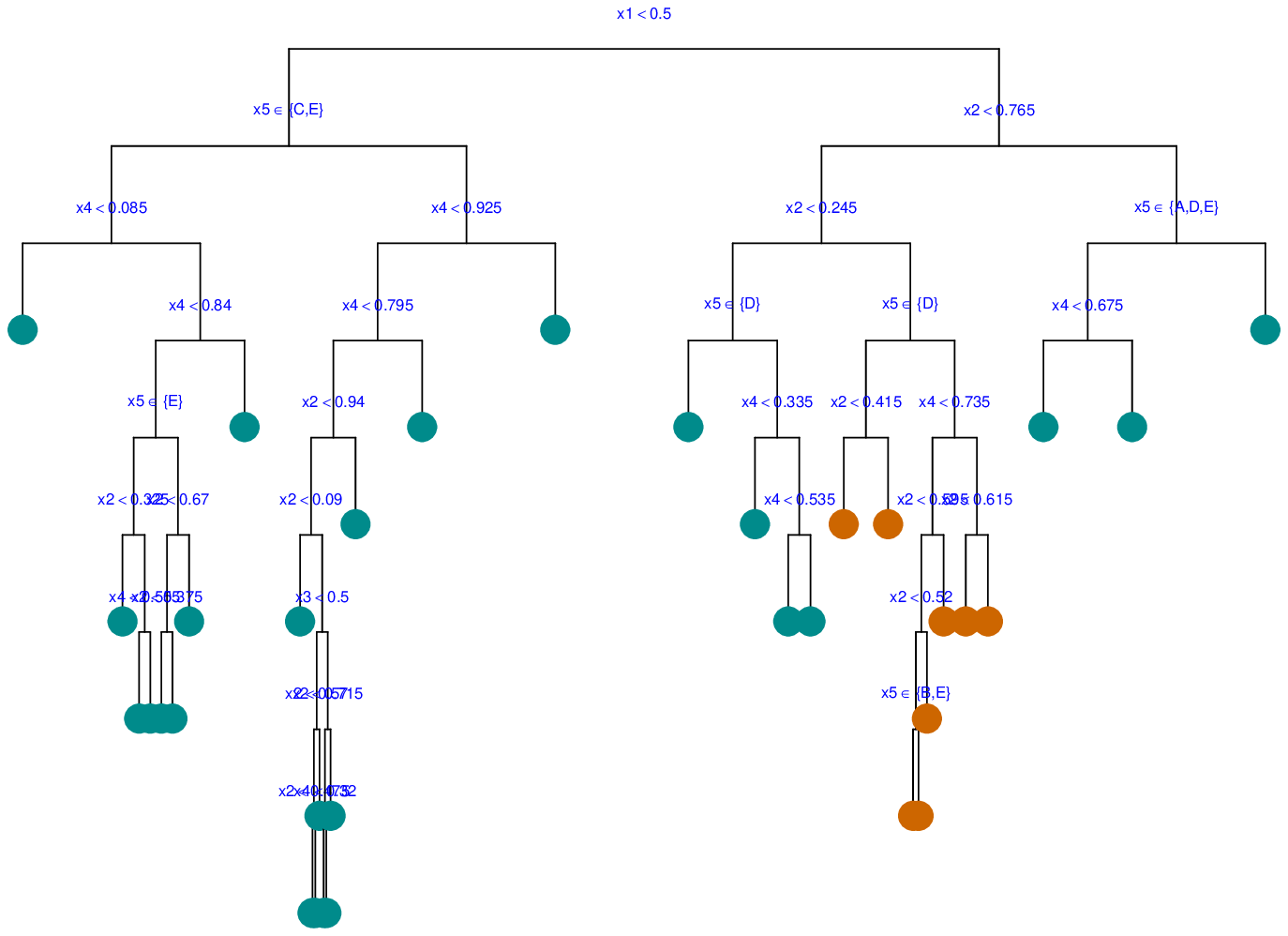}
         \caption{Initial Tree with Colored Leaves}
         \label{figIa}
     \end{subfigure}
\vspace{0.3in}

     \begin{subfigure}[b]{.65\textwidth}
         \centering
         \includegraphics[scale=.7,angle=270, width=\textwidth]{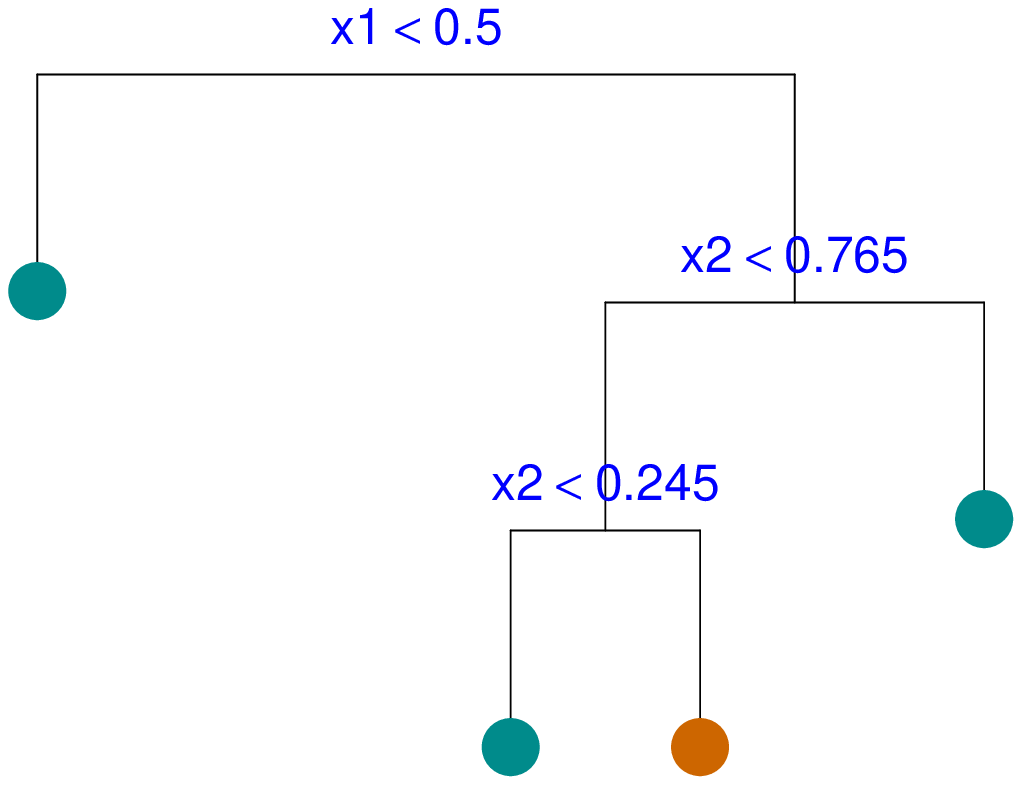}
         \caption{Final Tree after Shearing}
         \label{figIb}
     \end{subfigure}
        \caption{Illustration of Coloring and Shearing with data generated from Model C in Equation (3). (a) The initial large tree, where leaf nodes fall into two groups, represented by different colors. (b) The final tree structure after the shearing process. Tree shearing follows a top-down approach with a simple rule: prune any node whose descendants belong to the same color group. 
        \label{figI-NodeFusion}}
\end{figure}
\clearpage }

Panel~(a) displays the large initial tree constructed by SurvTreeFuL, along with the results of the coloring step obtained via fused regularization. A subsequent top-down \emph{shearing} step is then applied using a straightforward rule: prune any internal node whose descendant leaves all belong to the same color group. The result of this shearing process is the final SurvTreeFuL model, shown in Panel~(b). The main takeaway from this illustration is that once a grouping or coloring scheme for the terminal nodes is identified, the shearing step can be performed efficiently to derive the final tree structure. The fused regularization step facilitates this grouping process effectively. Together, they provide an efficient and principled alternative to traditional CART-style pruning for tree model or tree size selection.

Furthermore, our node fusion approach can be effectively combined with traditional pruning. A practically useful strategy is to first apply standard pruning to obtain a final tree, and then apply fused regularization to further merge its terminal nodes. This may lead to a more parsimonious and interpretable model. This two-step approach is justified by the extended tree model class $\mathcal{S}(\mathcal{T}_0)$ defined in Section~2.3, with the goal of identifying the best model within $\mathcal{S}(\mathcal{T}_0)$. A feasible implementation of this two-step procedure consists of: (1) obtaining a candidate tree structure from $\mathcal{T}_0$ using traditional pruning methods, and (2) applying node fusion via fused regularization to refine it. The first step does not require identifying the optimal tree structure; rather, it suffices to select a simpler tree. If a more rigorous selection is desired, validation techniques such as test-sample evaluation or $V$-fold cross-validation can be employed. These same validation methods can also be used in the second step to evaluate candidate models within $\mathcal{S}(\mathcal{T}_0)$. In practice, beginning the node fusion process from a simpler tree structure has proven advantageous. When the initial tree is overly large, it often includes leaves with inflated or deflated hazard estimates, which impairs the effectiveness of leaf sorting and reduces the chance of successful fusion. The proposed two-step approach mitigates this issue by simplifying the initial structure, thereby enhancing the performance of SurvTreeFuL in identifying meaningful groupings.

\renewcommand{\tabcolsep}{6.pt}
\renewcommand{\arraystretch}{1.5}
\renewcommand{\baselinestretch}{1.0}
\begin{table}[h]
\caption{Bootstrap Bias Correction (BBC) for Estimating Hazard Ratio and SD in Group Summary. Results are based on 100 simulation runs for each model. In each run, a training sample of size $n=600$ was generated, with a 50\% censoring rate applied.}
\vspace{.2in}
\centering
\begin{tabular}{lcccccrrcc} \hline \hline
	&		&	Test Data	&	\multicolumn{3}{c}{Uncorrected}					&&	\multicolumn{3}{c}{Bias-Corrected}					\\ 	 \cline{4-6} \cline{8-10}
Model	&	Estimate	&	Mean	&	Bias	&	MAD	&	MSE	&&	Bias	&	MAD	&	MSE	\\ 	\hline
Tree2	&	$\hat{\beta}$	&	3.210	&	0.253	&	0.682	&	0.154	&&	0.030	&	0.718	&	0.099	\\ 	
	&	SD	&	5.624	&	0.454	&	1.541	&	0.428	&&	0.076	&	1.327	&	0.228	\\ 	\hline
Linear	&	$\hat{\beta}$	&	1.880	&	0.414	&	1.283	&	0.223	&&	$-0.038$	&	1.107	&	0.063	\\ 	
	&	SD	&	5.894	&	0.407	&	1.193	&	0.265	&&	$-0.199$	&	1.158	&	0.244	\\ 	\hline
KAN	&	$\hat{\beta}$	&	1.195	&	0.416	&	0.823	&	0.217	&&	0.041	&	0.719	&	0.038	\\ 	
	&	SD	&	5.032	&	0.331	&	1.356	&	0.223	&&	$-0.077$	&	1.423	&	0.166	\\ 	\hline
\end{tabular}
\label{tbl-I}
\end{table}

\section{Additional Numerical Results}
\label{sec-AddResults} 

Referring back to Section~3.5, additional simulation studies were conducted to evaluate the performance of the bootstrap bias correction (BBC) method using data generated from a linear model and a nonlinear model, denoted as Model E and Model F, respectively, in Table~1. A simulation setup similar to that used for the tree model in Section~3.5 was adopted. Figure~\ref{figII-BBC} presents the results. Panels~(a) and (b) display the training- and test-based estimates of the regression coefficient $\beta$ and its standard deviation (SD), respectively, for the linear Cox model. For the training-based estimates, two versions are shown: the uncorrected estimates (in gray) and the bias-corrected estimates using the BBC method (in blue). Corresponding plots for the nonlinear KAN model appear in Panels~(c) and (d). From these plots, we observe that, relative to the test-based `honest' estimates, the training-based estimates exhibit an upward bias when left uncorrected. The BBC method appears effective in mitigating this bias. Because both the linear and nonlinear models are smooth, and tree-based models approximate such functions through piecewise-constant structures, no distinct clustering patterns emerge in the estimates, unlike those seen in Figure~4. 

Additional numerical summaries from the simulation studies described above are provided in Table~\ref{tbl-I}. Specifically, the table includes the mean of the test data based estimates of $\beta$, as well as its SD, across 100 simulation replicates. For both the uncorrected and bias-corrected estimates from training data, we report the average bias, median absolute deviation (MAD) from the median, and mean squared error (MSE). The MAD is used as a measure of variation to reduce potential redundancy among the reported statistics. As shown in Table~\ref{tbl-I}, the bootstrap bias correction (BBC) method substantially reduces the average bias. Notably, the corrected estimates exhibit mixed signs in the bias, indicating that the upward bias present in the uncorrected estimates has been largely mitigated. A similar pattern is evident in the MSE, where the bias-corrected estimates consistently show lower values than their uncorrected counterparts. When comparing results across the three models, the improvement from bias correction appears to be smallest for the linear model, possibly due to the challenges tree-based methods face in approximating linearity. As for the MAD, the comparison between uncorrected and corrected estimates yields mixed results, suggesting that the BBC method may have limited impact on variability. Overall, the BBC approach shows promise in reducing estimation bias for the purpose of group-level summaries in SurvTreeFuL, though its benefits may vary depending on the underlying data-generating mechanism.

\vspace{.3in}
\noindent

\afterpage{
\begin{figure}[H]
     \centering
\includegraphics[scale=0.7, angle=0]{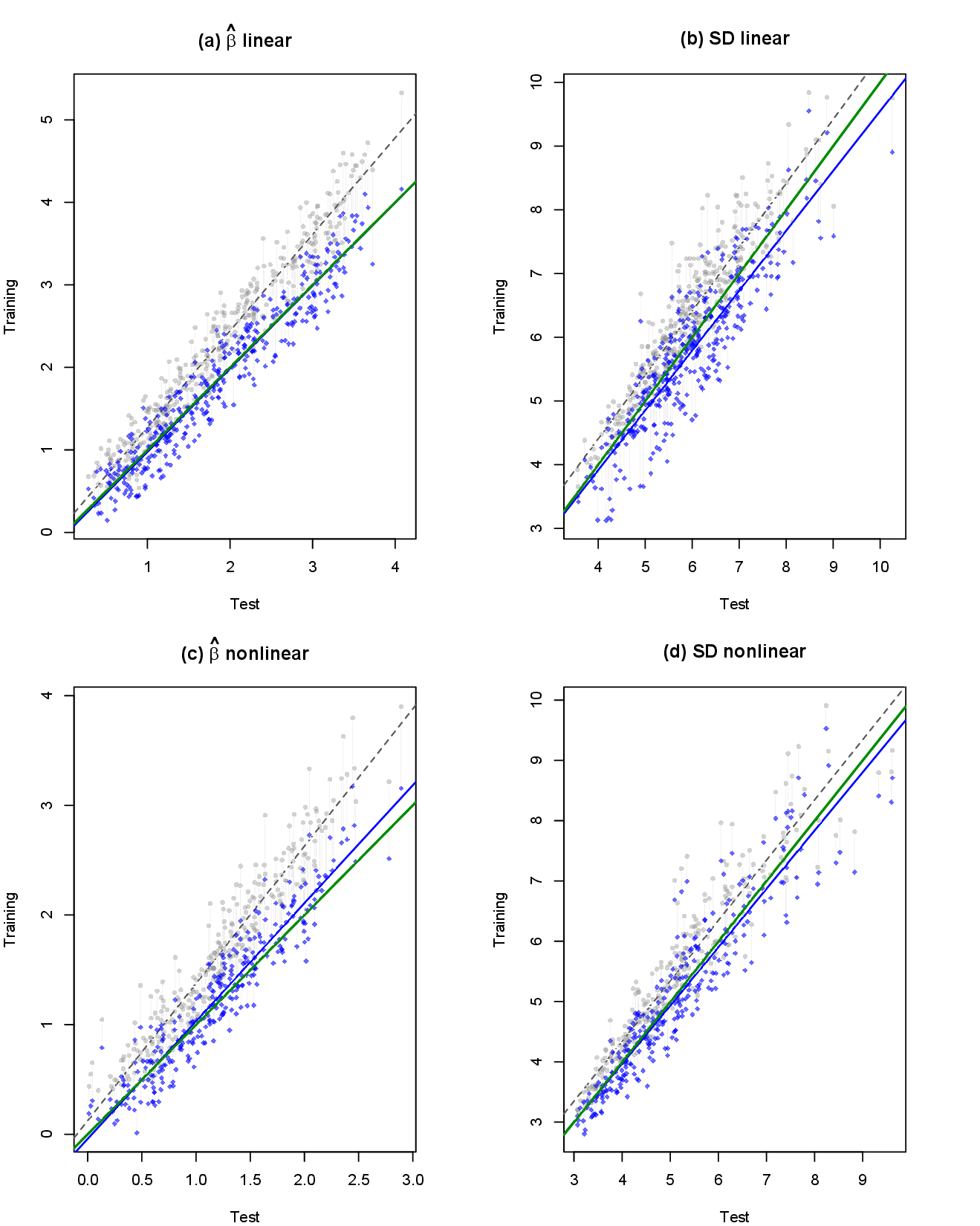}
\caption{Additional results for Cox model-based group summaries with bootstrap bias correction (BBC): (a, c) estimated coefficients $\hat{\beta}$ and (b, d) estimated standard deviations (SD), based on data generated from a linear Cox model (a, b) and a non-linear Cox model (c, d). Gray points and lines represent uncorrected estimates; blue points and lines represent BBC-adjusted estimates. 
\label{figII-BBC}}
\end{figure}
\clearpage }

\end{document}